\documentclass[a4paper,11pt]{article}
\usepackage{geometry}
\usepackage[svgnames]{xcolor}
\usepackage[colorlinks=true,urlcolor=black,linkcolor=blue,citecolor=blue,bookmarks=false]{hyperref}
\usepackage{amsfonts,amsmath,amssymb}
\usepackage{graphicx}
\usepackage{dsfont}
\usepackage{graphicx}
\usepackage{bm}
\usepackage[font=small,labelfont=bf]{caption}
\renewcommand*{\thepage}{\footnotesize\arabic{page}}
\usepackage{color}
\usepackage{enumitem}
\usepackage{boldline}
\usepackage{changepage}
\usepackage{cite}
\usepackage{textcomp}
\usepackage[title]{appendix}
\usepackage{url}
\usepackage[OT1]{fontenc}
 \usepackage{mathpazo}
\usepackage{authblk}
\usepackage{breakurl}
\usepackage{fancyhdr}
\usepackage[export]{adjustbox}
\usepackage[percent]{overpic}
\usepackage{floatrow}
\usepackage{multirow}
\usepackage{colortbl}

\title{\bf Identifying subdominant collective effects in a large motorway network}

\author{Shanshan Wang \thanks{shanshan.wang@uni-due.de}, Michael Schreckenberg and Thomas Guhr}
\affil{\textit{Faculty of Physics, University of Duisburg--Essen, Duisburg, Germany}}

\date{\today}
 
\begin{document}
\maketitle

\noindent {\bf Abstract.}
In a motorway network, correlations between parts or, more precisely, between the sections of (different) motorways, are of considerable interest. Knowledge of flows and velocities on individual motorways is not sufficient, rather, their correlations determine or reflect, respectively, the functionality of and the dynamics on the network. These correlations are time-dependent as the dynamics on the network is highly non-stationary. Apart from the conceptual importance, correlations are also indispensable to detect risks of failure in a traffic network. Here, we proceed with revealing a certain hierarchy of correlations in traffic networks that is due to the presence and to the extent of collectivity. In a previous study, we focused on the collectivity motion present in the entire traffic network, i.e.~the collectivity of the system as a whole. Here, we manage to subtract this dominant effect from the data and identify the subdominant collectivities which affect different, large parts of the traffic network.  To this end, we employ a spectral analysis of the correlation matrix for the whole system. We thereby extract information from the virtual network induced by the correlations and map it on the true topology, i.e.~on the real motorway network. The uncovered subdominant collectivities provide a new characterization of the traffic network. We carry out our study for the large motorway network of North Rhine-Westphalia (NRW), Germany.

\vspace{0.5cm}
\noindent{\bf Keywords\/}: complex system, traffic network, correlation matrix, spectral decomposition, $k$-means clustering, collectivities, strongly correlated groups
\vspace{1cm}

\noindent\rule{\textwidth}{1pt}
\vspace*{-1cm}
{\setlength{\parskip}{0pt plus 1pt} \tableofcontents}
\noindent\rule{\textwidth}{1pt}

\section{Introduction}
\label{sec1}

A traffic network as a complex system~\cite{May1972,Ladyman2013} is composed of all road sections on all links, i.e.~on all parts of the different motorways between ramps and crosses. These sections are connected with each other via the motion of the vehicles, resulting in highly non-stationary time series of traffic flows and velocities. The corresponding time series between different road sections are correlated~\cite{Wang2020}. Importantly, these sections can be, for example, neighboring on the same motorway or sections on different motorways far away from each other. Obviously, the correlations of the latter are of particular interest. Here, we recall that correlations do not necessarily reflect causalities, nevertheless the correlations are needed to assess the functionality of the network.

However, the information provided by the correlations between the many sections in a large traffic network has to be structured and condensed to be useful when trying to understand the functionality of the system. Here, we proceed with demonstrating that the correlations of a traffic network feature a certain hierarchy in the degree of collectivity. In a previous work~\cite{Wang2021}, we put forward a spectral analysis of the correlation matrices and showed that the largest eigenvalues correspond to the collective dynamics on the entire network. In the present study, we employ a recently developed, mathematically well-defined method~\cite{Heckens2020,Wang2020} to subtract this collectivity of the system as a whole from the data. This facilitates analyzing the subdominant collectivities which are present only in parts of the traffic network, not in the entire system. These subdominant collectivities are encoded in the large spectral outliers after the largest one has been removed. Importantly, there are only few such subdominant effects, rendering their identification a means to characterize the traffic network. We perform our analysis for the large motorway network of North Rhine-Westphalia (NRW), Germany.

Our approach extends a by now standard procedure in analyzing data from financial markets~\cite{Bouchaud2003,Liu1999,Gopikrishnan2001,Plerou2002}.  The largest eigenvalue of a correlation matrix describes the behavior of the financial market as a whole, while the subleading eigenvalues are due to the industrial sectors. The crucial difference to traffic is the necessity to identify the analogs of these industrial sectors, i.e.~the groups of strongly correlated sections. While the industrial sectors are a priori known for economic reasons, their traffic analogs cannot be inferred form additional input. Hence, it is a challenge and a major objective of this study to develop a method for the identification of the strongly correlated groups. This is related to the fact that we have to deal with only one network in finance, namely the virtual one induced by the correlations, while there is a second and true one in traffic, given by the real topology of the motorway map.

Identification of groups with strong correlations is of course an often encountered challenge in the study of many other complex systems, even though the detailed questions, methods and approaches differ.  For example, local perturbations, such as blackouts in power grids~\cite{Baldick2008}, congestions in traffic networks~\cite{Toroczkai2004,Daqing2014} and bankruptcies in credit networks~\cite{Caldarelli2013} can prompt cascading failures of the whole system. Related are the exploration of critical bottlenecks~\cite{Li2015} and critical phenomena~\cite{Zeng2019}, as well as the identification of metastable or quasi-stationary states~\cite{Zeng2020,Wang2020} in traffic networks.  We also notice that our study is different from previous ones~\cite{Li2015,Zeng2019,Zeng2020} focusing on urban networks. Our motorway network covers urban traffic and traffic in the countryside, thereby adding a new aspect highly relevant for territorial states.

The paper is organized as follows. In section~\ref{sec2}, we introduce our data set. In section~\ref{sec3}, we briefly sketch the construction of reduced-rank correlation matrices, compare the spectral information of the standard and reduced-rank correlation matrices, and develop an optimized $k$-means clustering method. In section~\ref{sec4}, we identify strongly correlated groups of motorway sections with empirical data, disclose their spectral and geographic features, and further figure out the relation between the dominant eigenvalues and geographic distributions of sections. We finally conclude our results in section~\ref{sec5}.

\section{Datasets}
\label{sec2}

\begin{figure}[pt]
\begin{center}
\includegraphics[width=0.9\textwidth]{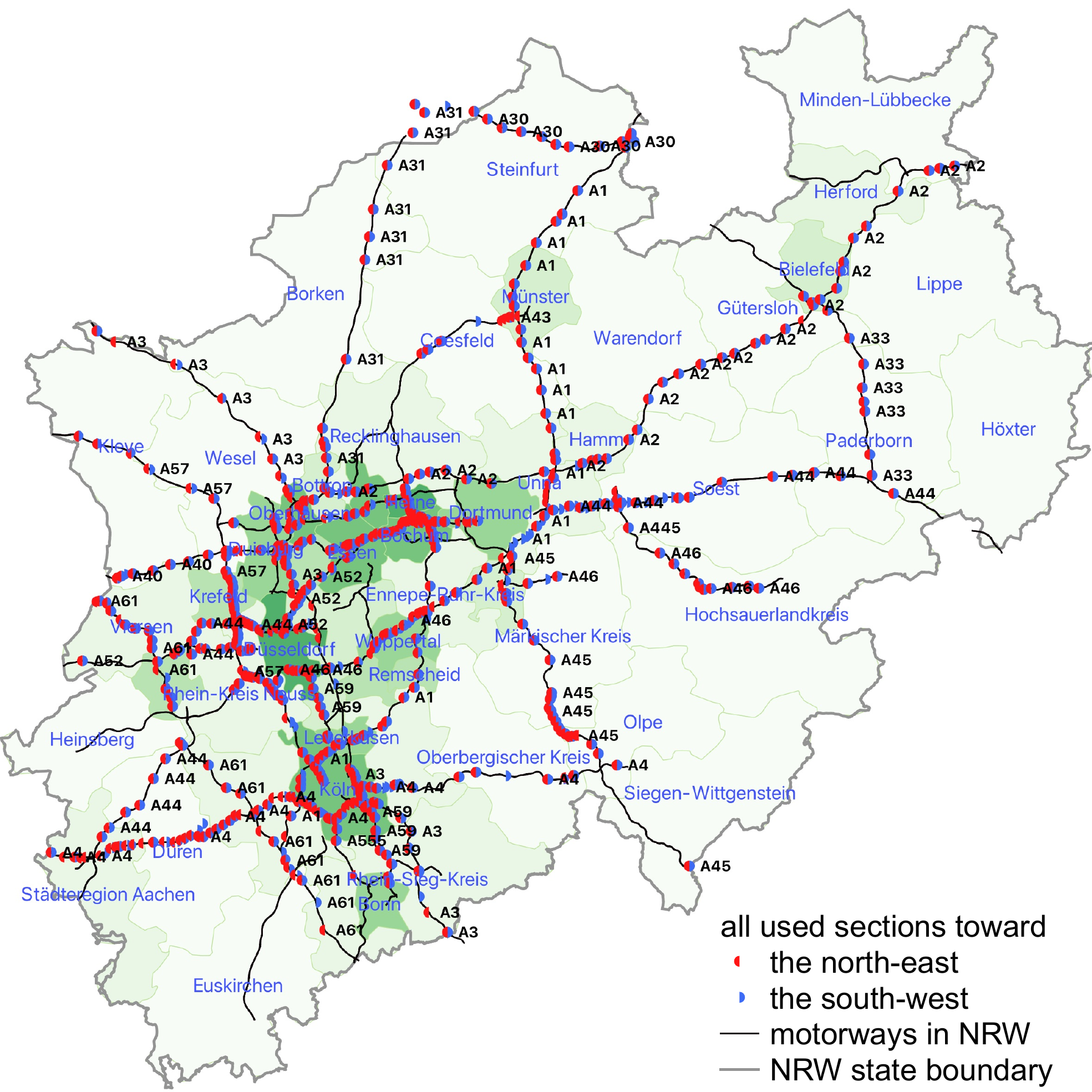}
\caption{The geographic distributions of 1179 available sections on 22 motorways of NRW in Germany. The green background indicates the population density of districts in NRW. The darker the green background, the higher the population density is. The data of administrative borders of districts (green lines) in NRW, licensed under BY-2.0, is provided by \copyright~GeoBasis-DE / BKG 2020~\cite{border,licence} and the data of population density in NRW, also licensed under BY-2.0, is provided by \copyright~Statistische \"{A}mter des Bundes und der L\"{a}nder, Germany~\cite{population,licence}. The data of motorways (black lines) and the outside administrative boundaries of NRW (grey lines), licensed under ODbL v1.0, is provided by \copyright~OpenStreetMap contributors~\cite{osmcopyright,osm}. The map is developed with QGIS 3.4~\cite{qgis}. }
\label{fig1}
\end{center}
\end{figure}

Our traffic data are collected by inductive loop detectors from $N=1179$ sections on 22 motorways in North Rhine-Westphalia (NRW), Germany, shown in figure~\ref{fig1}. The data with the resolution of one minute covers 80 discontinuous days, including 64 workdays and 16 holidays, in 2017. Here the weekends and public holidays of NRW in 2017 are all named as holidays. On each selected day, the ratio of missing values in the traffic data for each section is less than $60\%$. To be more specific, if considering the traffic data of each section at each day as a sub-dataset, $97.95\%$, $92.44\%$ and $59.65\%$ sub-datasets have less than $20\%$, $10\%$ and $1\%$ missing values, respectively. Regarding the data quality, the missing values are filled by their nearest non-missing values. We verified that the way of filling missing values has a negligible effect on our empirical results. This is important, as the missing values have a negative effect on the following spectrum decomposition. For each time, the data set includes the information of traffic flows and velocities for every lane on every section. The traffic flow gives the number of vehicles per unit time. Divided by the velocity, it yields the flow density that measures the number of vehicles per unit distance. For each section $n$ at each time $t$, we aggregate the traffic flows $q_{nl}(t)$ and the flow densities $\rho_{nl}(t)$ across all lanes $l$. The average velocity $v_n(t)$ for the same section at the same time can be obtained by
\begin{equation}
v_n(t)=\frac{\sum_lq_{nl}(t)}{\sum_l\rho_{nl}(t)} \ , \quad n=1,\cdots, N \ .
\label{eq2.3}
\end{equation}
Our study considers the case of all vehicles without distinguishing cars and trucks unless specific instructions, but distinguishes the cases of workdays and holidays due to different traffic behaviors.

 \section{Methods}
 \label{sec3}
 
To fix the notations and conventions, we begin with introducing the standard correlation matrix in section~\ref{sec31}. To subtract the dominant collectivity from the data, we remove, in a mathematically clean way, the largest eigenvalue from the standard correlation matrix in section \ref{sec32}. The resulting reduced-rank correlation matrix features the subdominant collectivities. We compare the spectral information between the standard and reduced-rank correlation matrices in section~\ref{sec33}. To identify the strongly correlated groups, we carry out a $k$-means clustering in a proper subspace of the eigenvectors in section~\ref{sec34}.

 \subsection{Standard correlation matrices}
  \label{sec31}
 
 For each section $n$, we have a time series of velocities $v_n(t)$ with the length $T$. The mean value and the standard deviation of this time series can be expressed by
 \begin{equation}
\mu_n=\frac{1}{T}\sum_{t=1}^{T} v_n(t)  \ , \quad n=1,\cdots, N \ , 
\label{eq3.1.4}
\end{equation}
and
\begin{equation}
\sigma_n= \sqrt{\frac{1}{T}\sum_{t=1}^{T}\big(v_n(t)-\mu_n\big)^2 }\ , \quad n=1,\cdots, N \ ,
\label{eq3.1.5}
\end{equation}
respectively. We normalize each element of this time series to zero mean and unit standard deviation by 
\begin{equation}
M_n(t)=\frac{v_n(t)-\mu_n}{\sigma_n} \ , \quad n=1,\cdots, N \ .
\label{eq3.1.6}
\end{equation}
Thus, we obtain a $N\times T$ data matrix $M$ whose $n$-th row is the normalized time series $M_n(t)$, $t=1,\cdots,T$. Therefore, the $N\times N$ correlation matrix of sections is given by
\begin{equation}
C=\frac{1}{T}MM^{\dag} \ .
\label{eq3.1.7}
\end{equation}
As explained in reference~\cite{Wang2021}, the largest eigenvalue of the correlation matrix $C$ captures the dominant effect, i.e.~the collective behavior of significant sections in the whole motorway network in NRW. For example, the rush hours contribute to this collectivity, but they are not the only reason for it, as the temporal analysis in reference~\cite{Wang2021} reveals. Collective behavior present in financial markets is identified similarly. The largest eigenvalue~\cite{Plerou2002,Pharasi2019} is proportional to the average of all elements in the correlation matrix~\cite{Plerou2002,Stepanov2015}. We notice that the systems in question, financial markets as well as traffic networks, are highly non-stationary, featuring different dynamics and correlation structures at different times. Hence, the dynamics of the largest eigenvalue may be viewed as a moving frame from which we now wish to assess the remaining dynamics, in particular the subdominant effects.

\subsection{Reduced-rank covariance and correlation matrices}
 \label{sec32}

To separate the subdominant collectivities in parts of the system from the dominant one in the entire system, we need to remove the effect of the largest eigenvalue from the correlation matrix. One possible method is a linear regression with empirical data to obtain the residuals as the new data~\cite{Gopikrishnan2001,Plerou2002}, which yields a new correlation matrix without the effect of the largest eigenvalue from the standard correlation matrix. Here we remove the effect of the largest eigenvalue by combining a singular value decomposition with the reconstruction of correlation matrices~\cite{Heckens2020}. The resulting correlation matrix is referred to as a reduced-rank correlation matrix, see reference~\cite{Goldstein1997} in another context.

We normalize each element of time series of $v_n(t)$ only to zero mean instead of to both zero mean and unit standard deviation 
\begin{equation}
A_{n}(t)=v_{n}(t)-\mu_n\ .
\label{eq3.2.1}
\end{equation}
The $N$ time series $A_{n}(t)$ form a new $N\times T$ data matrix $A$. We apply a singular value decomposition and expand $A$ in a sum of dyadic matrices,
\begin{equation}
A=\sum_{n=1}^{L}S_nU_nV^{\dag}_n \ ,
\label{eq3.2.2}
\end{equation}
where $L=\mathrm{min}(N,T)$ and the $S_n$ are the singular values. There are $N$ singular values for $N<T$ and $T$ for $T\leq N$. Furthermore, $U_n$ and $V_n$ are the corresponding left and right singular eigenvectors, respectively, where $U_n$ has $N$ and $V_n$ has $T$ components. Removing the largest eigenvalue from $A$ yields a new $N\times T$ data matrix,
\begin{equation}
\tilde{A}= \sum_{n=1}^{L-1}S_nU_nV^{\dag}_n \ .
\label{eq3.2.4}
\end{equation}
We introduce a $T$ dimensional unit column vector $e=(1,\cdots, 1)$ and a $P$ dimensional zero column vector $\emptyset_P=(0,\cdots, 0)$ with $P=N$ or $T$ such that 
\begin{equation}
Ae=\emptyset_N \qquad \mathrm{and} \qquad  A^{\dag}Ae=\emptyset_T \ .
\label{eq3.2.5}
\end{equation} 
The first of equation~(\ref{eq3.2.5}) is simply the normalization of all time series $A_n(t)$, $t=1,\cdots, T$ to zero means, written in a linear-algebra notation. From equations~(\ref{eq3.2.2}) and (\ref{eq3.2.5}) and due to the linear independence of the $U_n$, we find
 \begin{equation}
V^{\dag}_ne=0  \ ,
\label{eq3.2.6}
\end{equation}
for all $n$ in the case of $N<T$ due to the existence of $N$ non-zero singular values. In the case of $T\leq N$, there are $T-1$ non-zero singular values and one zero singular value, such that equation~(\ref{eq3.2.6}) is fulfilled for all $n$ except for the one when $S_n=0$. In any case, the following equation holds,
\begin{equation}
\tilde{A}e=\sum_{n=1}^{L-1}S_nU_nV^{\dag}_n e=\emptyset_N \ .
\label{eq3.2.7}
\end{equation}
Hence, all time series in $\tilde{A}$ are normalized to zero mean. The reduced-rank data matrix $\tilde{A}$ therefore yields a well-defined covariance matrix,
\begin{equation}
\tilde{\Sigma}=\frac{1}{T}\tilde{A}\tilde{A}^{\dag} \ ,
\label{eq3.2.8}
\end{equation}
named reduced-rank covariance matrix. The elements of each row in $\tilde{A}$ can be normalized to unit standard deviation by dividing out the standard deviation of the row, 
\begin{equation}
\tilde{B} =\tilde{\sigma}^{-1} \tilde{A} \ ,
\label{eq3.2.9}
\end{equation}
where $\tilde{\sigma}$ is the diagonal matrix of the standard deviations
\begin{equation}
\tilde{\sigma}=\mathrm{diag} \left(\tilde{\sigma}_1,\cdots,\tilde{\sigma}_N\right) \  \quad \mathrm{with}\quad  \tilde{\sigma}_n=\sqrt{\tilde{\Sigma}_{nn}} \ .
\label{eq3.2.10}
\end{equation} 
Utilizing the definition of a correlation matrix, we find the $N\times N$ reduced-rank correlation matrix  
\begin{equation}
\tilde{C}=\frac{1}{T}\tilde{B}\tilde{B}^{\dag}=\tilde{\sigma}^{-1} \tilde{\Sigma} \tilde{\sigma}^{-1} \ .
\label{eq3.2.11}
\end{equation}
For more details on the reduced-rank correlation matrix, we refer the reader to reference~\cite{Heckens2020}.

\subsection{Eigenvalues of covariance and correlation matrices}
 \label{sec33}
 
In the following, we compare the spectral information between the standard and reduced-rank correlation matrices. The spectral decomposition of the standard covariance matrix reads
\begin{equation} 
\Sigma=\frac{1}{T}AA^{\dag} =\sum_{n=1}^{L} \Lambda_n U_nU^{\dag}_n  \ ,
\label{eq3.3.1}
\end{equation}
where the eigenvalues
\begin{equation}
\Lambda_n=\frac{S^2_n}{T} \ 
\label{eq3.3.2}
\end{equation}
are directly related to the singular values. The standard covariance and correlation matrices $\Sigma$ and $C$, respectively, are related by 
\begin{equation}
\Sigma=\sigma C \sigma \ , \quad \sigma=\mathrm{diag} (\sigma_1,\cdots,\sigma_L)
\label{eq3.3.3}
\end{equation}
with the standard deviations $\sigma_l=\sqrt{\Sigma_{ll}}$. With equations~(\ref{eq3.3.2}) and (\ref{eq3.3.3}), we are able to expand equation~(\ref{eq3.2.11}) as
\begin{eqnarray}
\tilde{C}&=&\tilde{\sigma}^{-1} \left(\Sigma-\Lambda_LU_LU^{\dag}_L \right)\tilde{\sigma}^{-1}  =\tilde{\sigma}^{-1} \left(\sigma C \sigma-\frac{S^2_L}{T}U_LU^{\dag}_L \right)\tilde{\sigma}^{-1}   \ .
\label{eq3.3.4}
\end{eqnarray}
It is worth mentioning that there are in total $N$ eigenvalues either for $\Sigma$ or for $C$ due to the matrix dimensions $N\times N$. The number of their non-zero eigenvalues is $L$ in the case of $N<T$ and $L-1$ in the case of $T\leq N$~\cite{Chakraborti2020,Heckens2020} with $L=\mathrm{min}(N,T)$. To avoid confusion, we list the numbers of non-zero eigenvalues in table~\ref{tab1}. In an ascending order, $\Lambda_L$ and $S_L$ are the largest non-zero eigenvalue and the largest non-zero singular value of $\Sigma$ and $A$, respectively. 
We further decompose the two correlation matrices in the above equation by
\begin{equation}
C=\sum\limits_{n=1}^{L}\lambda_n u_nu^{\dag}_n \ 
\label{eq3.3.5}
\end{equation}
and 
\begin{equation}
\tilde{C}=\sum\limits_{n=1}^{L}\tilde{\lambda}_n \tilde{u}_n\tilde{u}^{\dag}_n \ ,
\label{eq3.3.6}
\end{equation}
where we only consider the largest $L$ eigenvalues $\lambda_n$ and $\tilde{\lambda}_n$ of $C$ and $\tilde{C}$, respectively, and ignore the other zero eigenvalues. Besides, $u_n$ and $\tilde{u}_n$ are the corresponding $N$-component eigenvectors of $C$ and $\tilde{C}$, respectively. As there are $N-1$ non-zero eigenvalues for $N<T$ and $T-2$ non-zero eigenvalues for $T\leq N$ as shown in table~\ref{tab1}, the minimal eigenvalue is always zero, i.e., $\tilde{\lambda}_1=0$.

\begin{table}[b]
\begin{footnotesize}
\begin{center}
\caption{The numbers of non-zero eigenvalues of matrices for different cases}
\begin{tabular*}{\textwidth}{c@{\extracolsep{\fill}}ccccccc}
\hlineB{2}
matrix & dimensions  & \multicolumn{2}{c}{$N<T$} & &\multicolumn{2}{c}{$T\leq N$}\\
\cline{3-4} \cline{6-7}
		&		& number of & number of & &  number of  &  number of  \\
		&		& all eigenvalues & non-zero eigenvalues & & all eigenvalues& non-zero eigenvalues\\
\hline
$A$ 	   	& $N\times T$ 	&$N$  &  $N$ 	&  	&  $T$  &  $T-1$\\
$\Sigma$ & $N\times N$ 	&$N$  &  $N$	&	& $N$  &  $T-1$\\
$C$ 	   	& $N\times N$ 	&$N$  &  $N$	&	& $N$  &  $T-1$\\
$\tilde{C}$ & $N\times N$	&$N$  &  $N-1$	&	& $N$  &  $T-2$\\
\hlineB{2}
\label{tab1}
\end{tabular*}
\end{center}
\end{footnotesize}
\vspace*{-0.5cm}
\end{table}%

\subsection{Identifying the strongly correlated groups by clustering}
\label{sec34}

The reduced-rank correlation matrix is free of the dominant collectivity.  However, as mentioned in the introduction, we face a situation different from finance, because we have to identify the analogs of the industrial sectors without additional input. Put differently, while the obvious economic relations tell us, how to choose a proper basis that clearly reveals the industrial sectors, we now have to find such a basis for the traffic system. Luckily, it turns out that this task can be solved in a relatively small subspace spanned by only few eigenvectors corresponding to the few large eigenvalues of the reduced-rank correlation matrix. The reordering of the basis can then be done by a clustering algorithm. Equation~(\ref{eq3.3.6}) is equivalent to
\begin{equation}
\tilde{C}=\tilde{u}\tilde{\lambda} \tilde{u}^{\dag} \ , \quad \tilde{\lambda}=\mathrm{diag}(\tilde{\lambda}_1,\cdots,\tilde{\lambda}_N) 
\label{eq3.4.1}
\end{equation}
with the eigenvalues $\tilde{\lambda}_1,\cdots,\tilde{\lambda}_N$ in an ascending order, and $\tilde{u}$ is a $N\times N$ orthogonal matrix whose columns are the corresponding eigenvectors $\tilde{u}_n$, such that
\begin{equation}
\tilde{u}=\big[\tilde{u}_1  \cdots  \tilde{u}_N  \big] \ .
\label{eq3.4.2}
\end{equation}
Either for $N<T$ or for $T\leq N$, the matrix $\tilde{C}$ always has $N$ eigenvalues in total, as listed in table~\ref{tab1}, and the $N$ corresponding eigenvectors. To reduce the noise and to lower the dimension to the relevant one for clustering, we focus on the large eigenvalues. To find $k$ correlated groups of sections, we use the eigenvector information corresponding to the largest $k-1$ eigenvalues~\cite{Ding2004} and define the $N\times(k-1)$ eigenvector matrix
\begin{equation}
\tilde{u}^{(k-1)}=\big[\tilde{u}_{N-(k-1)+1}  \cdots  \tilde{u}_N \big] \ 
\label{eq3.4.3}
\end{equation}
as our data for $k$-means clustering~\cite{Lloyd1982,Forgy1965}. It then follows that $k$ is the number of clusters (or groups) in $k$-means clustering. 

To determine the number $k$, we resort to the Marchenko-Pastur eigenvalue density~\cite{Marchenko1967} as a qualitative guideline. The Marchenko-Pastur distribution, resulting from the spectral density of a correlation matrix,
\begin{equation}
\rho(\lambda)=\sum\limits_{n=1}^{N}\delta(\lambda-\lambda_n) \ ,
\label{eq3.4.4}
\end{equation}
is the large-$N$ eigenvalue density for a fully random correlation matrix. It is known~\cite{Laloux2000,Plerou2000,Plerou2002} that it also describes the bulk of many large non-random correlation matrices. The bulk of eigenvalues is between $\lambda_{-}$ and $\lambda_{+}$
\begin{equation}
\lambda_{\pm}=1+\frac{N}{T}\pm2\sqrt{\frac{N}{T}} \ 
\label{eq3.4.5}
\end{equation}
with $T\neq N$. Typically, there are large eigenvalues outside the bulk, often way outside the bulk. They indicate strongly correlated groups~\cite{Gopikrishnan2001,Plerou2002}, and their number $k-1$ is the one we use for the clustering.

To carry out the clustering, we consider the $n$-th row of the matrix $\tilde{u}^{(k-1)}$, i.e., a vector, as an observation corresponding to section $n$. Therefore, clustering all these observations means clustering our sections. The components in this vector are the features for comparing the similarity of two observations. Here, we define the distance between two observations $i$ and $j$ as the squared Euclidean distance,
\begin{equation}
d_{ij}=\sum\limits_{m=N-k+2}^{N}(\tilde{u}_{im}-\tilde{u}_{jm})^2 \ ,
\label{eq3.4.6}
\end{equation}
which entries in a distance matrix $d$. Employing the distance matrix, we implement $k$-means clustering~\cite{Lloyd1982,Forgy1965} for our observations. The $k$-means clustering mainly contains the following steps:
\begin{itemize}[leftmargin=0.6cm]
\setlength\itemsep{-0.3em}
\item[(a)] Select $k$ initial centroids for observations. 
\item[(b)] Compute distances from each observation to every centroid.
\item[(c)] Assign each observation to the cluster with the closest centroid.
\item[(d)] Recalculate the average of the observations assigned to each cluster in order to find a new centroid for each cluster. 
\item[(e)] Repeat steps (b)--(d) until the assignments of observations do not change or iterations reach the preset maximal number. 
\end{itemize}
The silhouette value quantifies how similar an observation is to its own cluster as compared to other clusters~\cite{Kaufman2009}. It ranges from -1 to +1, where a high positive value indicates an appropriate classification of an observation under its own cluster, while a low or negative value indicates a poor clustering configuration. To optimize our clustering, we carry out the whole procedure of clustering as follows:
\begin{itemize}[leftmargin=0.6cm]
\setlength\itemsep{-0.3em}
\item[(1)] Perform $k$-means clustering with the squared Euclidean distance.
\item[(2)] Refine the clustering as follows:
\begin{itemize}[leftmargin=1cm]
\item[(2.1)] Validate the consistency within clusters by silhouette values and reassign all observations with negative silhouette values to an additional cluster, i.e., $(k+1)$-th cluster.
\item[(2.2)] Reassign each observation in $(k+1)$-th cluster separately to all $k+1$ clusters and calculate the silhouette value of that observation for every assignment. 
\item[(2.3)] Reassign that observation to the cluster in which the observation acquires the maximal silhouette value comparing to in other clusters.
\item[(2.4)] Calculate silhouette values of all observations and reassign the observations with negative silhouette values to $(k+1)$-th cluster.
\item[(2.5)] Repeat steps (2.2)--(2.4) until  the assignments of observations do not change or iterations reach the preset maximal number.
\end{itemize}
\item[(3)] Validate the consistency within clusters by silhouette values.
\item[(4)] Reorder the indices of $k+1$ clusters according to the contribution of eigenvectors.
\end{itemize}

\section{Empirical results}
\label{sec4}
 
We analyze empirical data of the motorway network in NRW, Germany. Using the methods sketched in section~\ref{sec3}, we identify in section~\ref{sec41} the subdominant collectivities, i.e.~the strongly correlated groups of motorway sections. We then analyze spectral properties and determine the relevant eigenvalues for each group in section~\ref{sec42}. In section~\ref{sec43}, we visualize and interpret geographic features of the identified correlated groups and relate them to traffic phases in Kerner's theory~\cite{Kerner2012}. We associate the relevant eigenvalues with geographic locations of the motorway sections in section~\ref{sec43}.

\subsection{Strongly correlated groups in the motorway network}
\label{sec41}

We work out the $1179\times 1179$ reduced-rank correlation matrices for workdays and holidays with the method described in section~\ref{sec32}, where $T=1440$. The largest eigenvalue of the standard correlation matrix, i.e., $\lambda_\mathrm{max}=135.6$ for workdays and $\lambda_\mathrm{max}=112.8$ for holidays, respectively, are subtracted applying the described procedure. According to our previous experience, the resulting reduced-rank correlation matrix is free of the dominant collective behavior that affects the system as a whole, see references~\cite{Heckens2020} for financial markets and~\cite{Wang2021} for the NRW motorway network. In the case of workdays, however, there is a deviation from the mentioned previous analyses. The numerical value $\lambda_\mathrm{2nd~max}=131.3$ of the second largest eigenvalue is rather close to that of the largest. Usually, the numerical value of the second largest eigenvalue is considerably smaller. Obviously, the second largest eigenvalue in the case in question reflects the presence of another collectivity affecting a large part of the system as well. Nevertheless, the distribution of the eigenvector components corresponding to the largest and the second largest eigenvalue worked out in reference~\cite{Wang2021} show differences for the significant participants, i.e., the significant motorway sections. As this indicates that also the acting mechanisms are different, we decided to proceed as in the previous analyses and as for the holidays in the present one. Hence, we only subtract the largest eigenvalue of the standard correlation matrix. Anticipating the later discussion, we mention that the strongly correlated groups related to the second largest eigenvalue for workdays, i.e.~to the largest eigenvalue of the reduced-rank correlation matrix, comprise a large part of the system, but not the system as a whole. This justifies applying the same procedure of analysis as previously, in particular for workdays and holidays.

\begin{figure}[tbp]
\begin{center}
\includegraphics[width=0.81\textwidth]{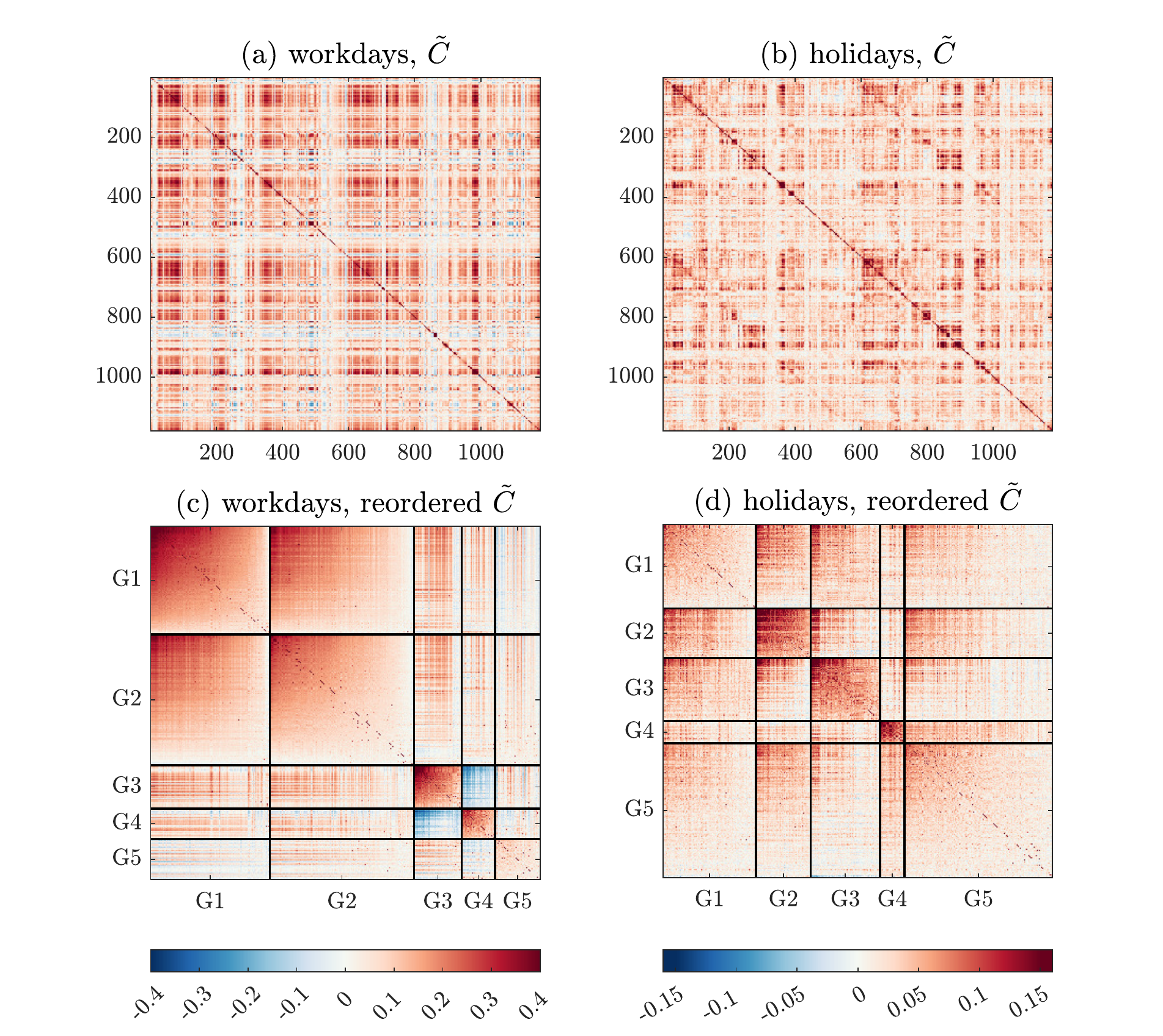}
\caption{The reduced-rank correlation matrices $\tilde{C}$ for workdays (a) and holidays (b), and the reduced-rank correlation matrices $\tilde{C}$ with reordered rows and columns for workdays (c) and holidays (d), where the color indicates the value of correlations and the black lines distinguish group 1 (G1), group 2 (G2), group 3 (G3), group 4 (G4) and group 5 (G5), respectively.}
\label{fig2}
\end{center}
\vspace*{-0.5cm}
\end{figure}

\begin{figure}[tbp]
\begin{center}
\includegraphics[width=0.78\textwidth]{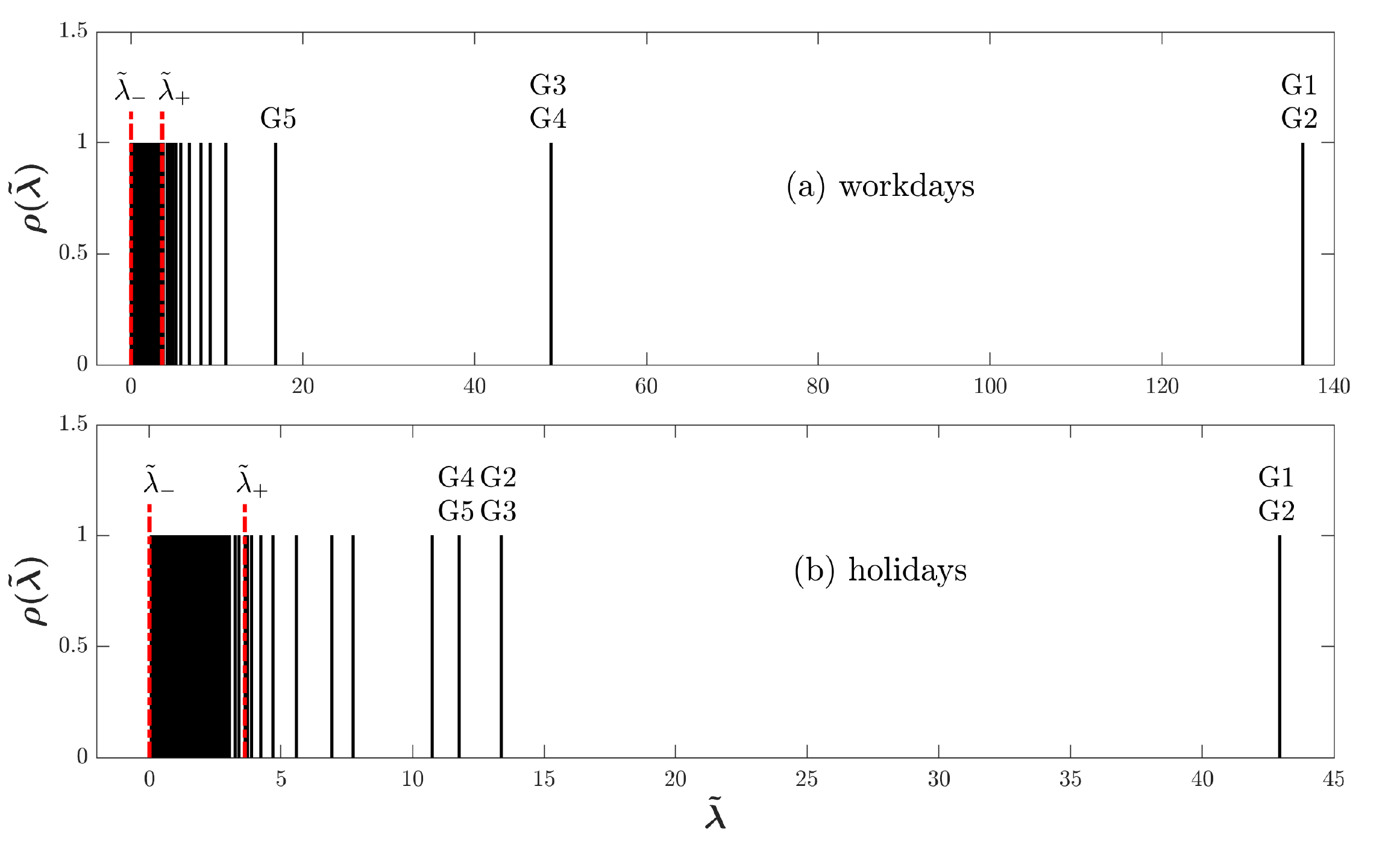}
\vspace*{-0.5cm}
\caption{The distributions of spectral density for the reduced-rank correlation matrices $\tilde{C}$, where $\tilde{\lambda}_{+}$ and $\tilde{\lambda}_{-}$ are the maximal and minimal Marchenko-Pastur eigenvalues, respectively. G$i$ stands for group $i$ with $i=1$, 2, 3, 4 and 5, respectively, and is marked near the corresponding relevant eigenvalue. }
\label{fig3}
\end{center}
\end{figure}

Removing the largest eigenvalues from the standard correlation matrices also reduces the strength of the remaining correlations among the sections, but makes the structures of correlation matrices more distinct, as shown in figure~\ref{fig2}. However, compared to the case of finance, the structures are not as strongly developed in the reduced-rank correlation matrices $\tilde{C}$. To identify correlated groups of sections, we apply the clustering method as described in section~\ref{sec34} to the eigenvectors of $\tilde{C}$. An important step is to determine the number $k$ of clusters. If $k$ is too small, the group information remains hidden, if $k$ is too large, it is blurred. The spectral density~(\ref{eq3.4.4}) of $\tilde{C}$, displayed in figure~\ref{fig3}, gives us a basic idea on how the large eigenvalues outside the bulk are distributed. It is worth mentioning that the eigenvalues, from the second smallest to the largest, of each reduced-rank correlation matrix $\tilde{C}$ correspond one-to-one to the eigenvalues, from the smallest to the second largest, of the standard correlation matrix $C$, but their numerical values are a bit shifted due to the necessary change in the normalization of $\tilde{C}$. For a figure of the spectral density of $C$, we refer the reader to reference~\cite{Wang2021}. The first three and the first four largest eigenvalues of the reduced-rank correlation matrices for workdays and for holidays, respectively, are much larger than $\tilde{\lambda}_{+}$ of equation (\ref{eq3.4.5}).  As the third and fourth eigenvalues for the case of holidays are close to each other, we use the first three eigenvalues for both cases such that $k-1=3$, i.e., $k=4$.

\begin{figure}[tbp]
\begin{center}
\includegraphics[width=0.9\textwidth]{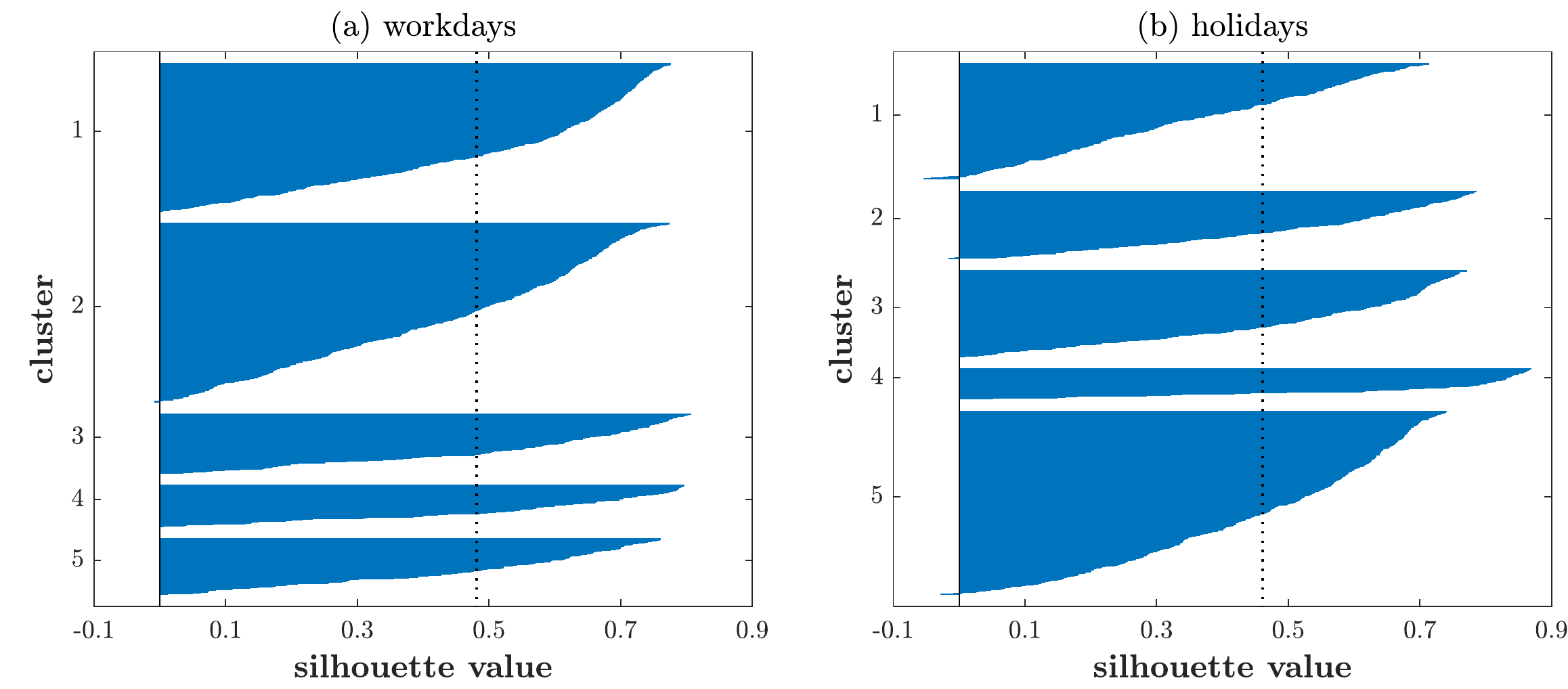}
\caption{The silhouette values for validating the consistency within five clusters. The dot line indicates the averaged silhouette value for workdays (a) and for holidays (b), respectively. }
\label{fig4}
\end{center}
\vspace*{-0.2cm}
\end{figure}

We carry out the clustering and find five well-classified correlated groups of sections, which can be validated by the silhouette values as depicted in figure~\ref{fig4}. The average silhouette values for both cases are close to 0.5, suggesting successful classifications in correlated groups for all sections. To visualize the correlation strength of each group, we reorder all rows and columns of the reduced-rank correlation matrices according to the indices of groups. The orders of rows and columns are always the same, implying that the diagonal elements represent the self-correlation and are equal to one. As a result, the sections within the same group are put together and organized in a descending order of correlations from top to bottom and from left to right in each group. The diagonal blocks of the reordered matrices in figure~\ref{fig2} reveal the internal correlations of each group. We can find at least three strongly correlated groups, for instance, the first four groups for workdays and the middle three groups for holidays. In particular, groups 3 and 4 are obviously anti-correlated for workdays.

\subsection{Spectral features of the strongly correlated groups}
\label{sec42}

\begin{figure}[tbp]
\begin{center}
\includegraphics[width=1\textwidth]{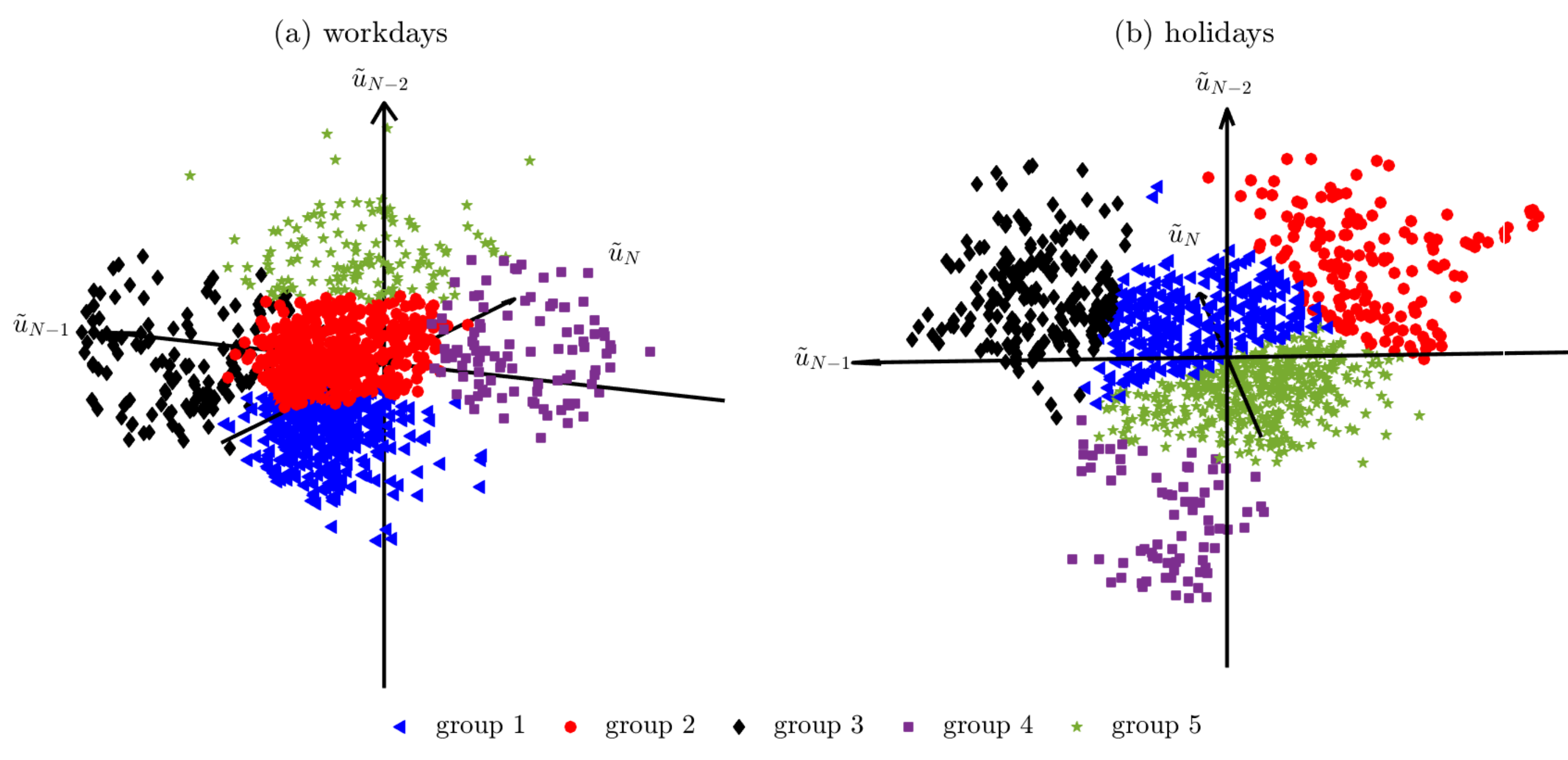}
\vspace*{-0.5cm}
\caption{The three-dimensional scatter plots of the eigenvector components corresponding to the largest three eigenvalues of the reduced-rank correlation matrices for workdays (a) and holidays (b), respectively. For a better angle of view, the axes are rotated so that some of axes may not be obvious.}
\label{fig5}
\end{center}
\end{figure}

\begin{figure}[tbp]
\begin{center}
\includegraphics[width=1\textwidth]{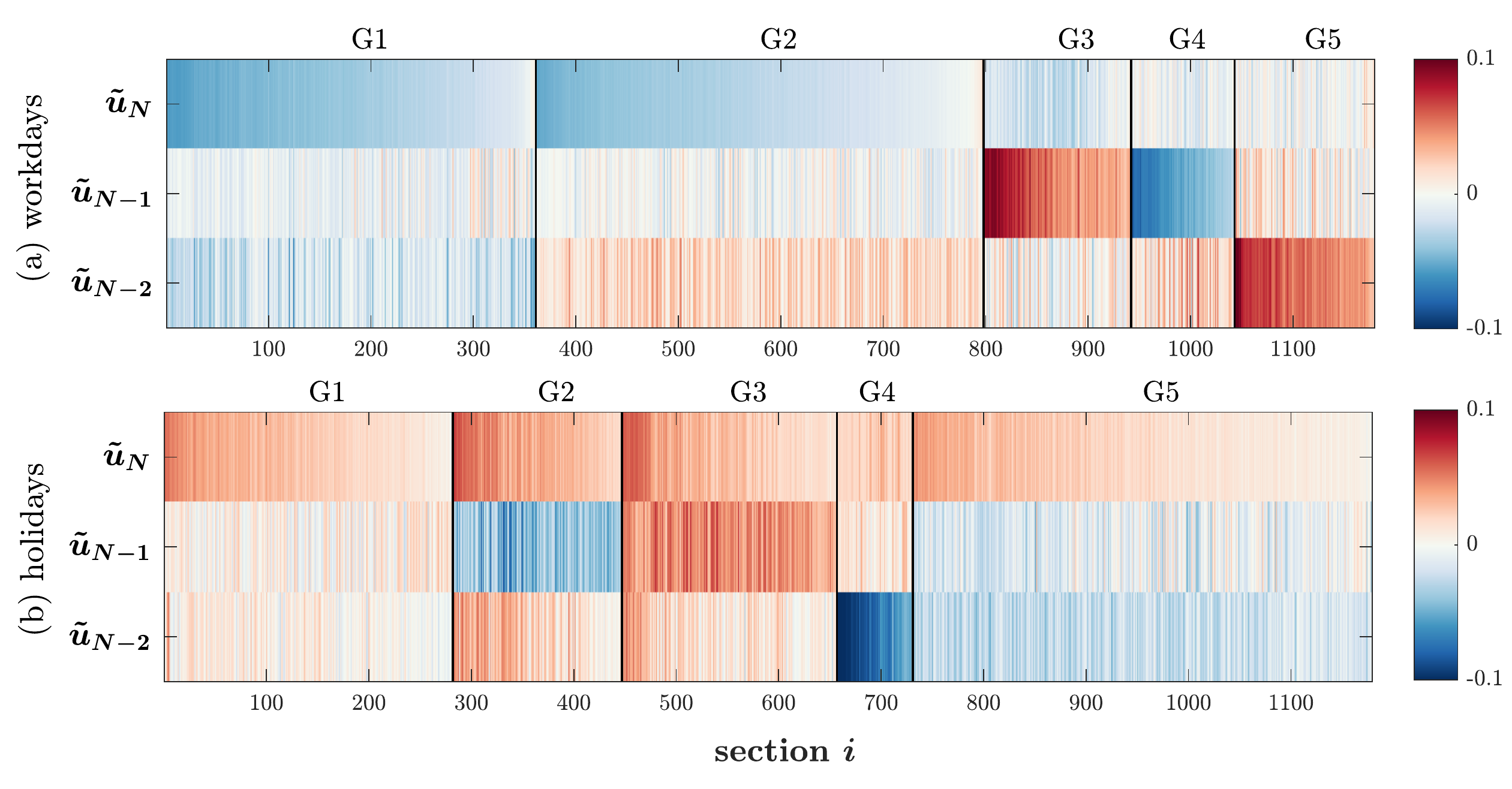}
\vspace*{-1cm}
\caption{The transposed eigenvector matrices~(\ref{eq3.4.3}), i.e., $(\tilde{u}^{(k-1)})^{\dag}$, for the largest three eigenvalues, where the eigenvector components along the horizontal axis are ordered corresponding to the reduced-rank correlation matrices in figure~\ref{fig2} (c) and (d). The black lines distinguish five groups (G1 -- G5). The color indicates the value of eigenvector components, i.e., $\tilde{u}_{iN}$, $\tilde{u}_{iN-1}$ and $\tilde{u}_{iN-2}$ in eigenvectors $\tilde{u}_{N}$, $\tilde{u}_{N-1}$ and $\tilde{u}_{N-2}$, respectively.}
\label{fig6}
\end{center}
\end{figure}

\begin{table}[b]
\begin{footnotesize}
\begin{center}
\caption{ The numbers of sections, the average correlation strength and the values of $\gamma^{(3)}_{j}$ in each group}
\begin{tabular*}{\textwidth}{c@{\extracolsep{\fill}}lrrrrcrrrrr}
\hlineB{2}
	&\multicolumn{5}{c}{for workdays} & & \multicolumn{5}{c}{for holidays} \\
\cline{2-6}  \cline{8-12}
 	 &$q$ & $\langle\tilde{C}_{ij}\rangle_{ij}$ &$\gamma^{(3)}_{1}$ &$\gamma^{(3)}_{2}$ & $\gamma^{(3)}_{3}$ &  & $q$ &$\langle\tilde{C}_{ij}\rangle_{ij}$ &$\gamma^{(3)}_{1}$ &$\gamma^{(3)}_{2}$&$\gamma^{(3)}_{3}$\\
\hline
 group 1	& 361   & \color{blue}  0.1789   &\color{red} 0.0136    & -0.0101  & -0.0035    &  & 282       & 0.036   & \color{red} 0.0097     &-0.0034   & -0.0063 \\
group 2	& 437   & \color{blue}  0.1048   &\color{red} 0.0086     & -0.0084   & -0.0003  &  & 165    &\color{blue} 0.0928    & \color{red} 0.0039 & \color{red} 0.0061   & -0.0100 \\
group 3	& 144   & \color{blue}  0.2062   & -0.0113  & \color{red}  0.0256      & -0.0142  &  & 210      &\color{blue} 0.0721      & -0.0006    & \color{red} 0.0139      & -0.0133 \\
group 4	& 101   & \color{blue}  0.1784   & -0.0179   & \color{red}  0.0241      & -0.0063 &  & 74       &\color{blue} 0.1088         & -0.0145     	& -0.0236   & \color{red} 0.0380 \\
group 5	& 136   &  0.0639  & -0.0175  & -0.0115  & \color{red} 0.0290  &  &448  & 0.0262  & 0.0008  & -0.0037      & \color{red} 0.0030 \\
\hlineB{2}
\label{tab2}
\end{tabular*}
\end{center}
\end{footnotesize}
\end{table}%

Since the clustering is based on the principal spectral information, we wish to explore how the three largest eigenvalues contribute to each correlated group of sections. Figure~\ref{fig5} displays three-dimensional scatter plots of their eigenvector components. The groups are well separated from each other, in accordance with figure~\ref{fig4}. Each group in its three-dimensional eigenvector space is located mainly along one or two eigenvectors. We also visualize the eigenvector components of the largest three eigenvalues in figure~\ref{fig6}. For each group, at least one eigenvector is strongly occupied by either the positive or the negative components. We then extract the eigenvector matrix~(\ref{eq3.4.3}) of the largest three eigenvalues for each group and rebuild the index for section $i$ from 1 to the total number $q$ of sections in each group, listed in table~\ref{tab2}. With regard to the absolute eigenvector components, we can quantify the relative importance of the three eigenvalues by
\begin{equation}
\gamma^{(3)}_{j}=\frac{1}{q}\sum_{i=1}^{q}\left(|\tilde{u}_{iN-j+1}^{(\mathrm{G}g)}|-\frac{1}{3}\sum_{j=1}^{3}|\tilde{u}_{iN-j+1}^{(\mathrm{G}g)}|\right) \ 
\label{eq4.1.3}
\end{equation}
with $j=1$, 2 and 3 for the first, the second and the third largest eigenvalues, respectively. The superscript $(\mathrm{G}g)$ stands for group $g$ with $g = 1$, 2, 3, 4 and 5, respectively. A positive value of $\gamma^{(3)}_{j}$ means that the effect of the $j$-th largest eigenvalue on a group is more than the average effect of the three largest eigenvalues. As a result, the $j$-th largest eigenvalue is relevant in this group. 

\begin{figure}[tbp]
\begin{center}
\includegraphics[width=1\textwidth]{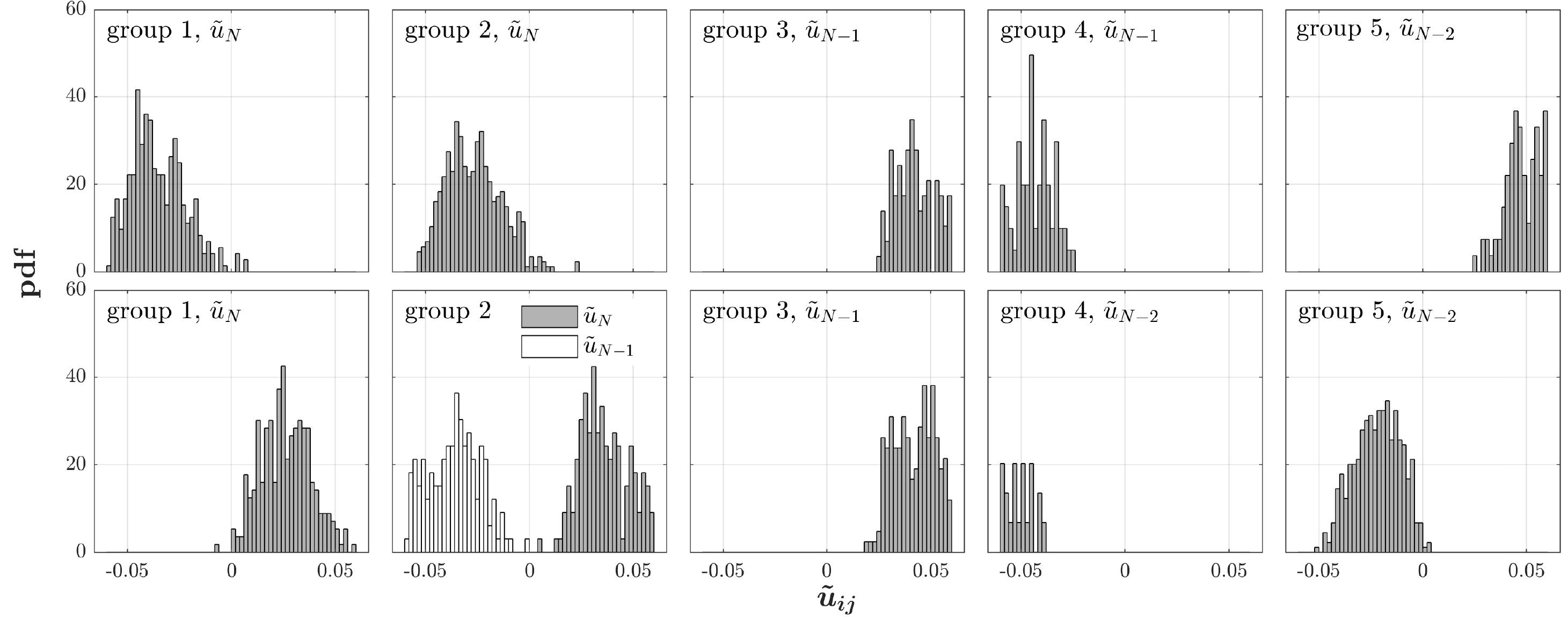}
\vspace*{-0.5cm}
\caption{The distributions of eigenvector components corresponding to the relevant eigenvalues of each group for workdays (upper row) and for holidays (bottom row), respectively.}
\label{fig7}
\end{center}
\vspace*{-0.5cm}
\end{figure}

According to $\gamma^{(3)}_{j}$, we identify the relevant eigenvalues for each group, as listed in table~\ref{tab2} and marked in figure~\ref{fig3}. For the case of workdays, groups 1 and 2 are mainly due to the effect of the largest eigenvalue, while group 5 is mostly due to the effect of the third largest eigenvalue. Interestingly, groups 3 and 4 are anti-correlated with each other, as shown in figure~\ref{fig2}, and very likely result from the opposite effects of the second largest eigenvalue, revealed by figure~\ref{fig6}. We notice that the strongly correlated groups 1 and 2 related to the largest eigenvalue of the reduced-rank correlation matrix comprise, in the case of workdays, a large part of the system, but not the system as a whole. This is a justification, as mentioned before, for our construction of the reduced-rank correlation matrix by subtracting the largest eigenvalue (of the standard correlation matrix) only, although the largest two eigenvalues (of the standard correlation matrix) have similar numerical values in the case of workdays. For the case of holidays, groups 1 and 3 are induced by the effects of the first and the second largest eigenvalues, respectively. In between is group 2 which is governed by the largest two eigenvalues. Regardless of the remarkable difference in the importance $\gamma^{(3)}_{j}$, groups 4 and 5 are strongly influenced by the third largest eigenvalues. The aforementioned findings are also supported by the distributions of eigenvector components corresponding to the relevant eigenvalues, as shown in figure~\ref{fig7}. The distributions mainly located on either the positive or the negative side indicate that almost all sections in each group are driven by the same effect represented by the relevant eigenvalue. In particular, the second largest eigenvalue governs both groups 3 and 4 for workdays, but the corresponding eigenvector components of the two groups are located on opposite sides around zero. This suggests that the opposite effects from the second largest eigenvalue work on the two groups. In addition, group 2 for holidays is governed by the largest two eigenvalues.

 \subsection{Geographic features of the strongly correlated groups}
\label{sec43}

\begin{figure}[tbp]
\begin{center}
\includegraphics[width=0.48\textwidth]{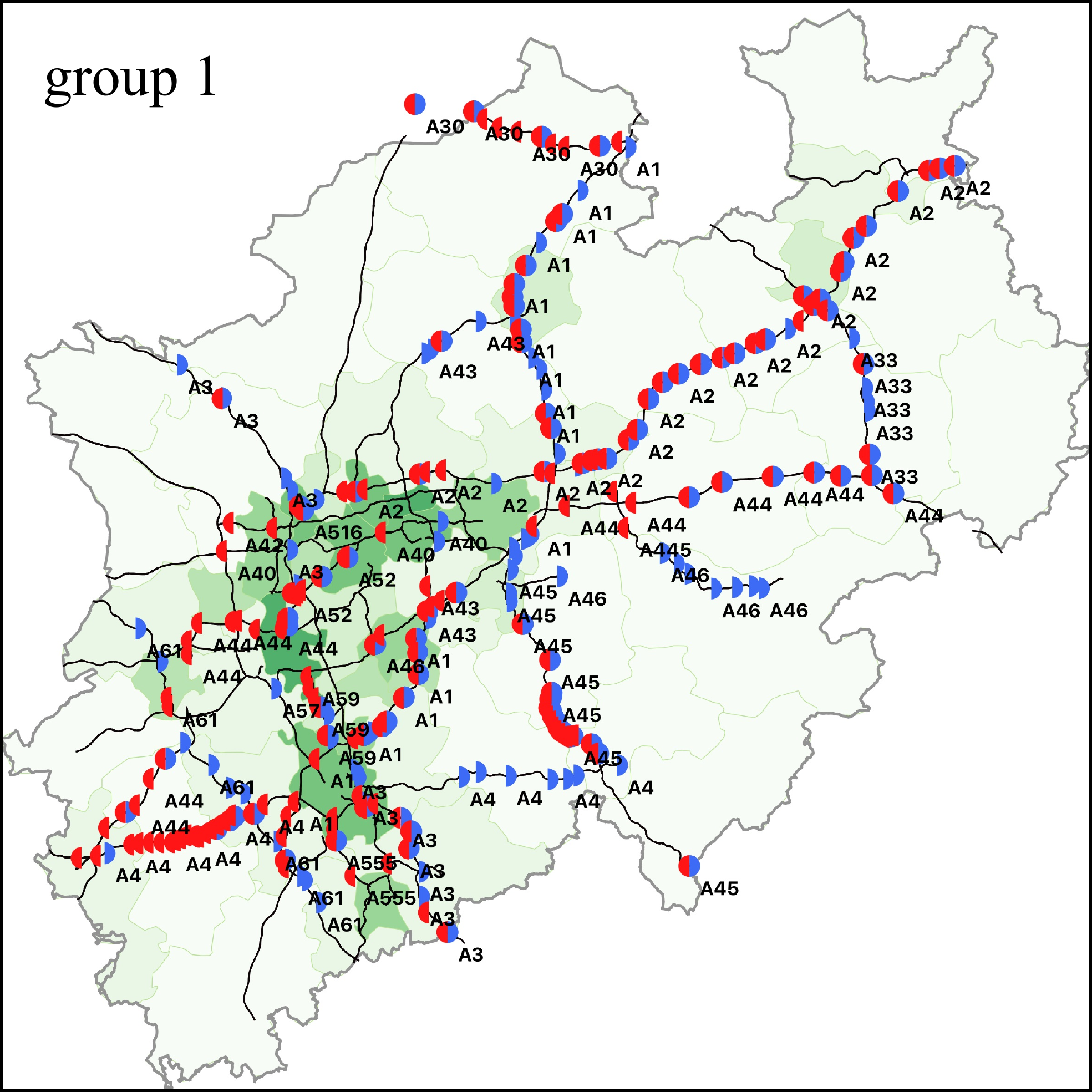}
\includegraphics[width=0.48\textwidth]{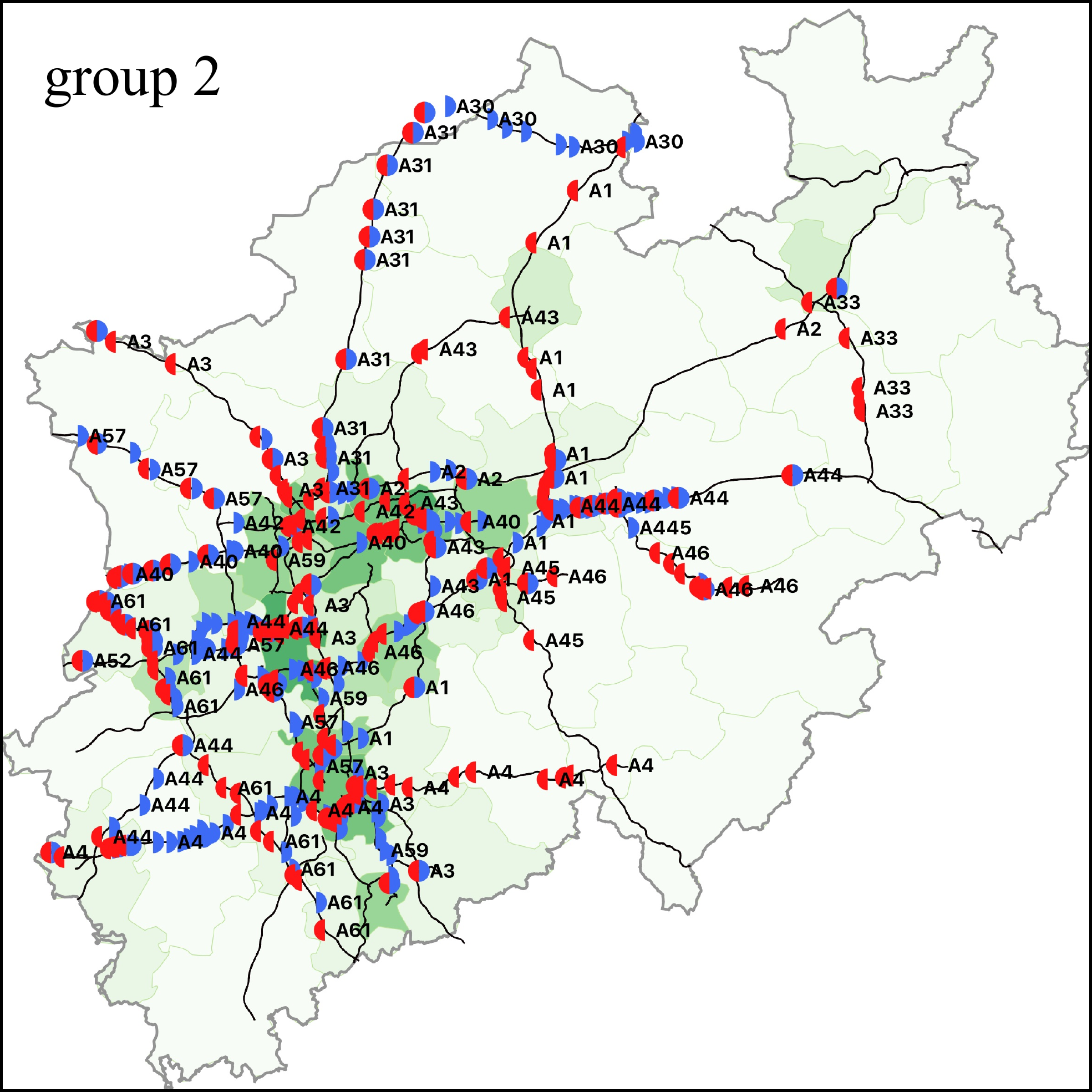}\\ \vspace*{0.07cm}
\includegraphics[width=0.48\textwidth]{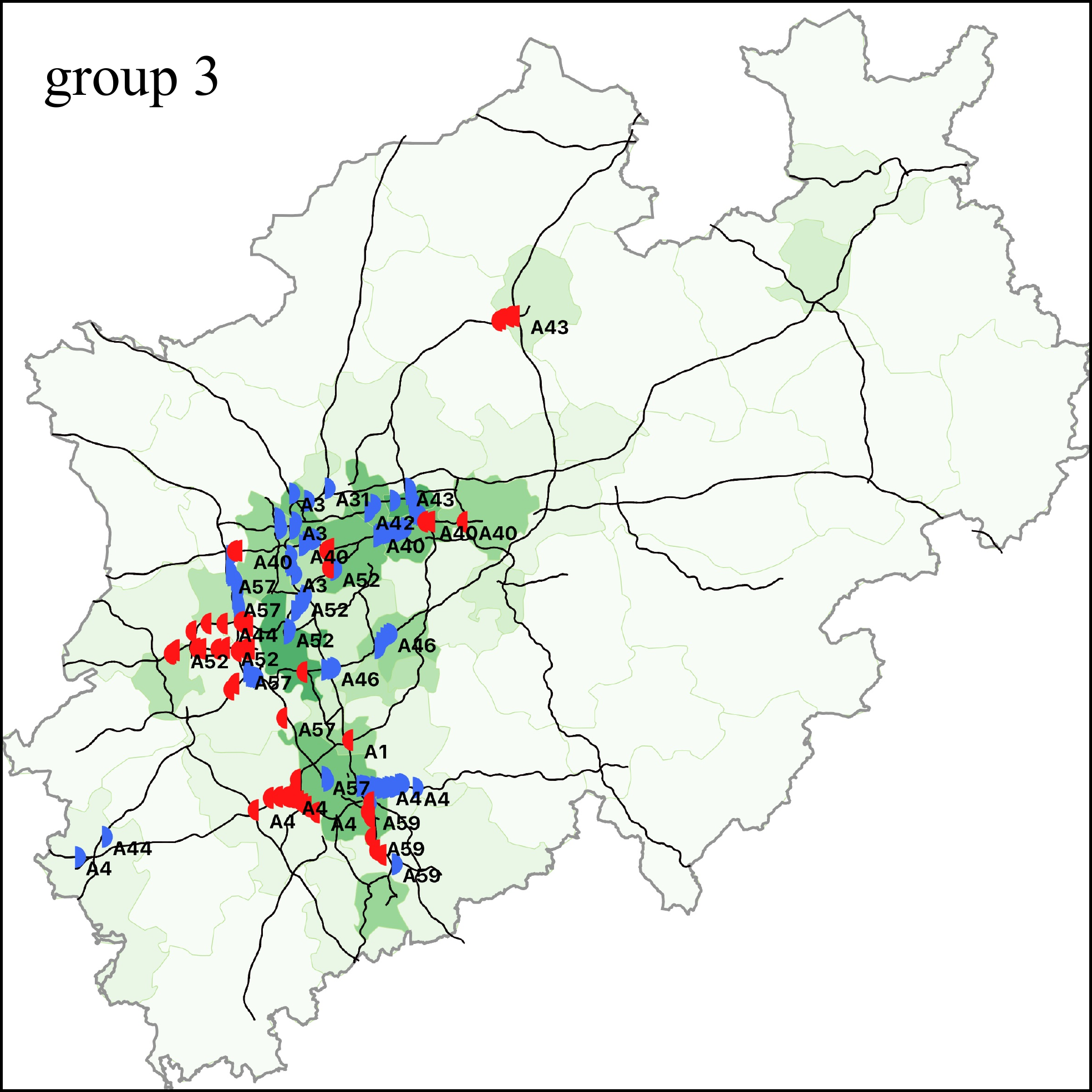} 
\includegraphics[width=0.48\textwidth]{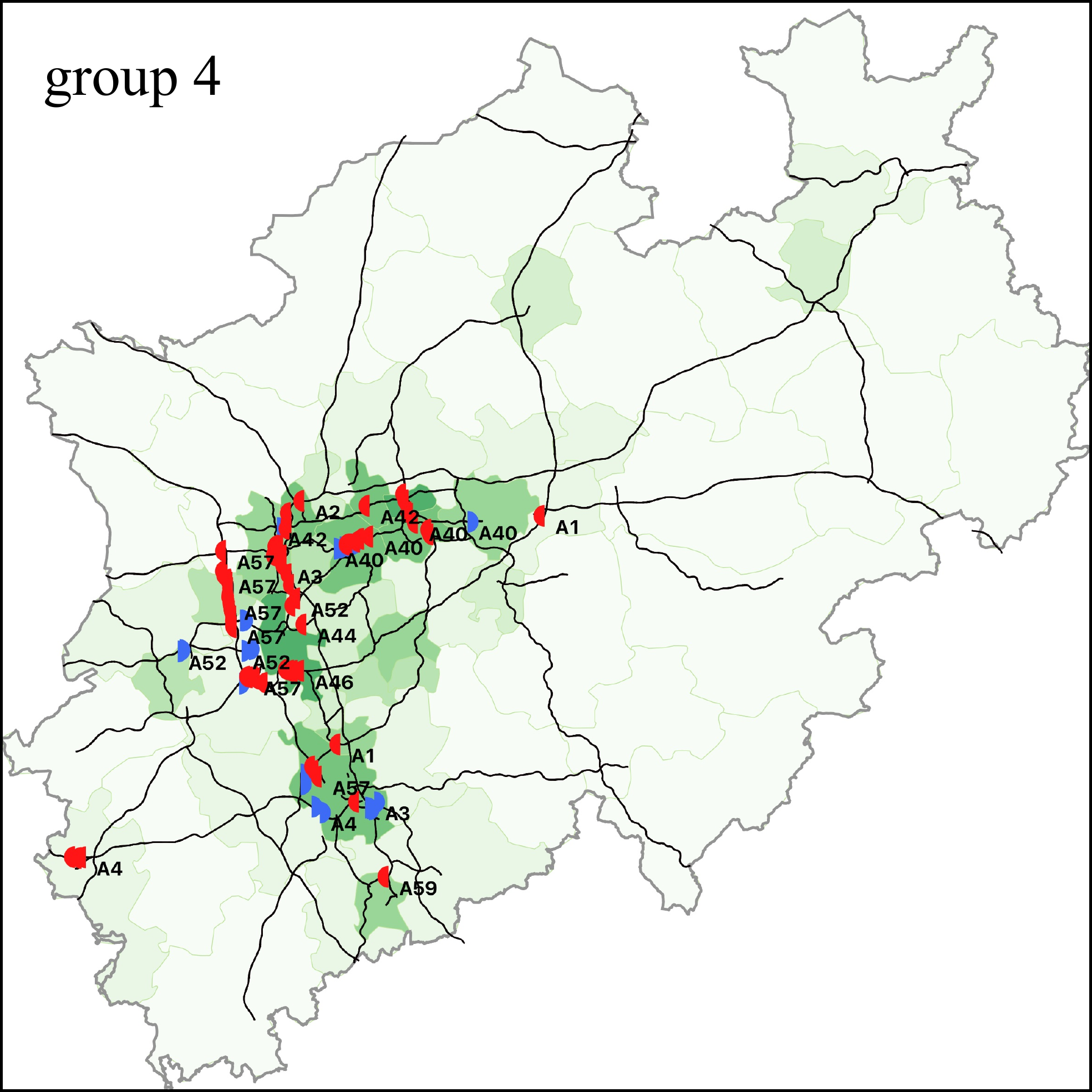} \\ \vspace*{0.07cm}
\floatbox[{\capbeside\thisfloatsetup{capbesideposition={right,center},capbesidewidth=0.465\textwidth}}]{figure}[\FBwidth]
{\caption{Five groups of motorway sections of NRW in Germany for workdays. The green background indicates the population density of districts in NRW. For the information and the data source of the base map, see figure~\ref{fig1}.}
\label{fig8}
}
{\includegraphics[width=0.48\textwidth]{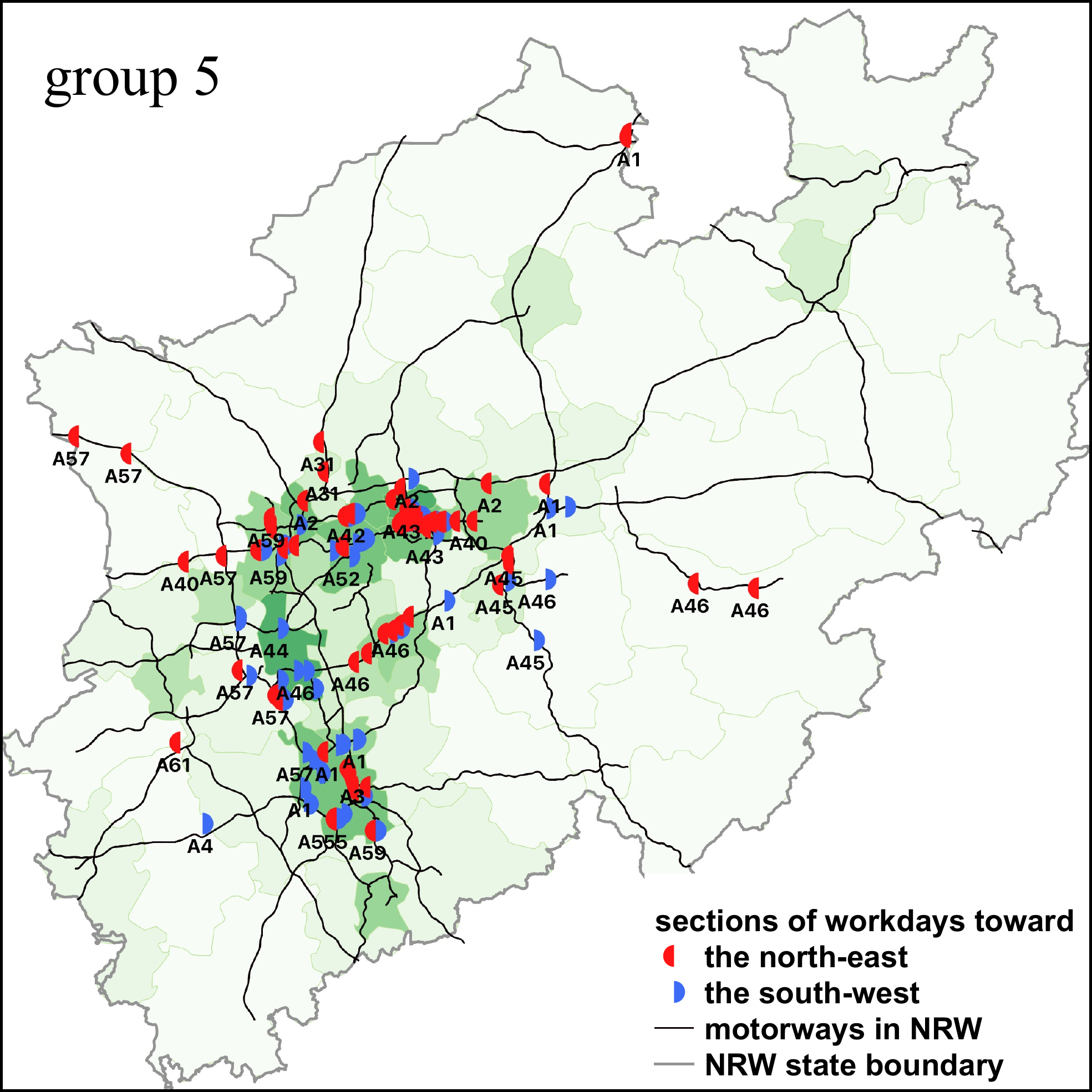}}

\end{center}
\end{figure}

\begin{figure}[tbp]
\begin{center}
\includegraphics[width=0.48\textwidth]{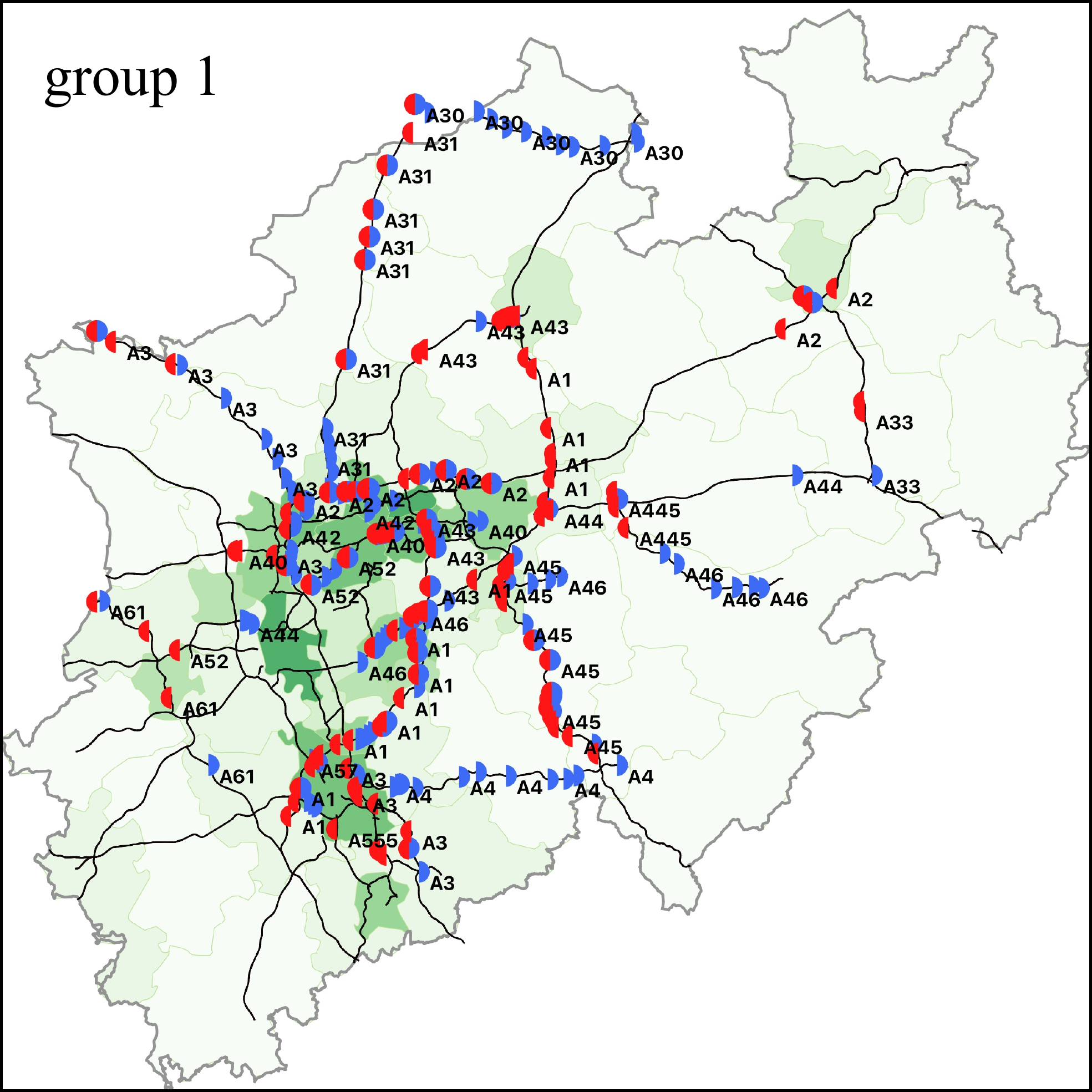} 
\includegraphics[width=0.48\textwidth]{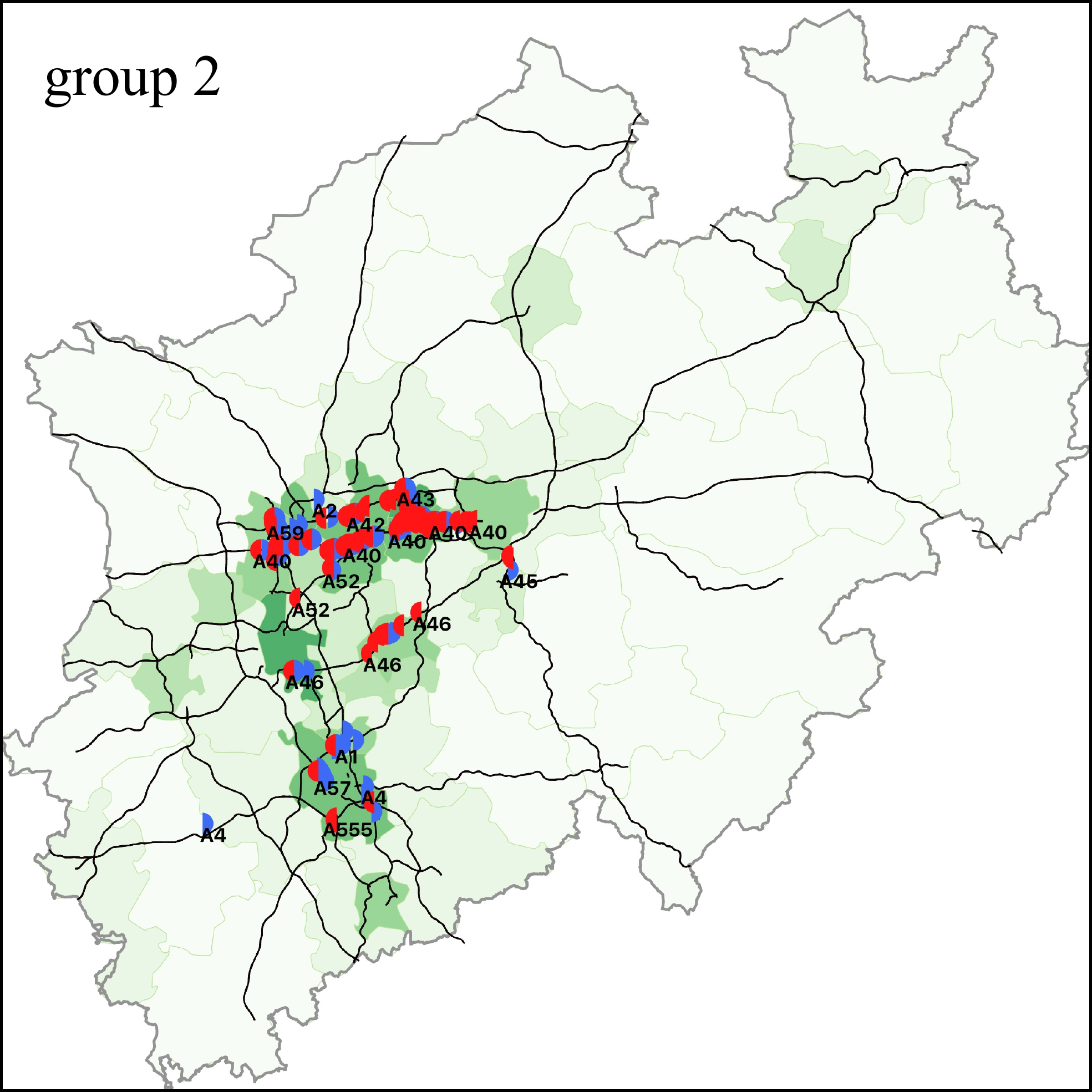} \\ \vspace*{0.07cm}
\includegraphics[width=0.48\textwidth]{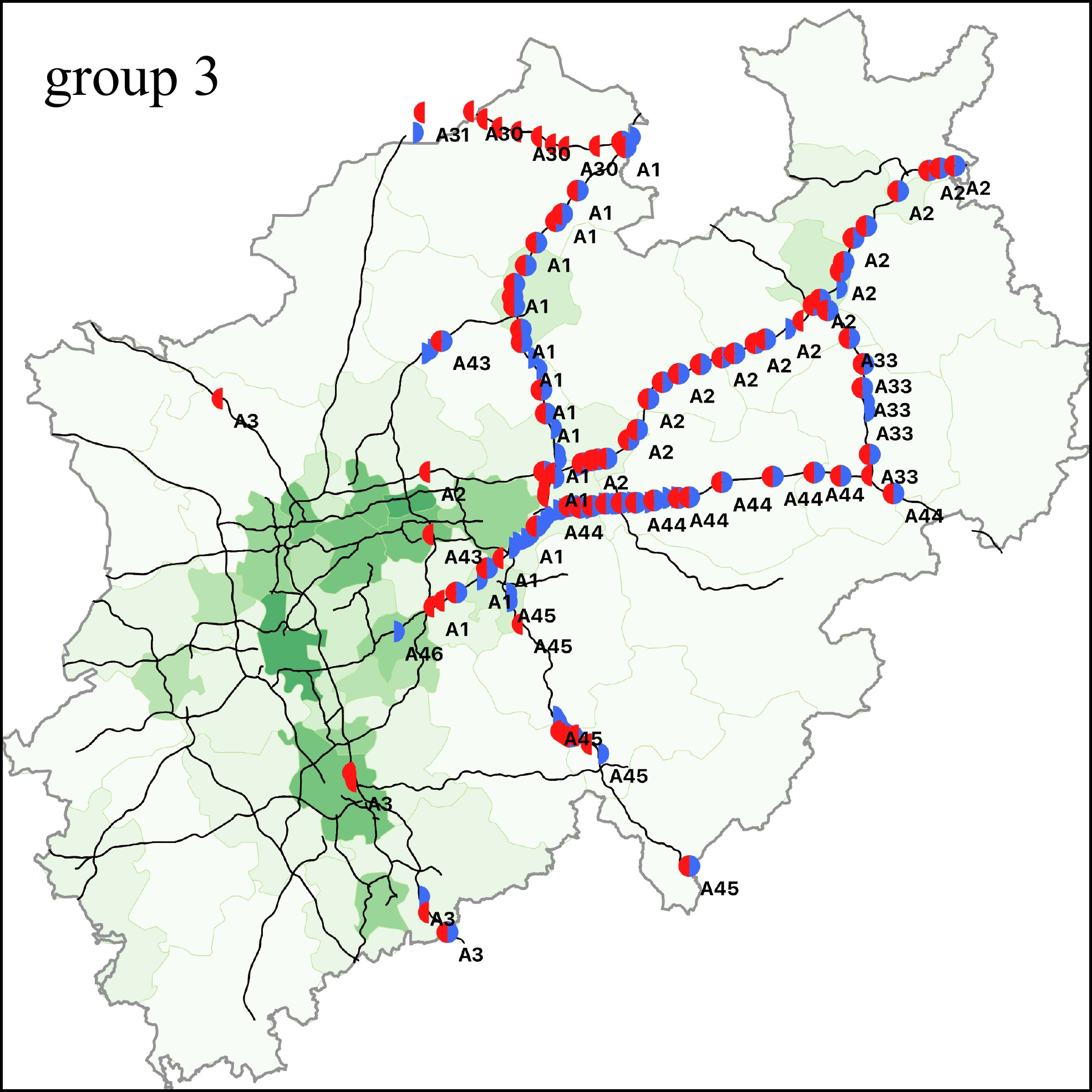} 
\includegraphics[width=0.48\textwidth]{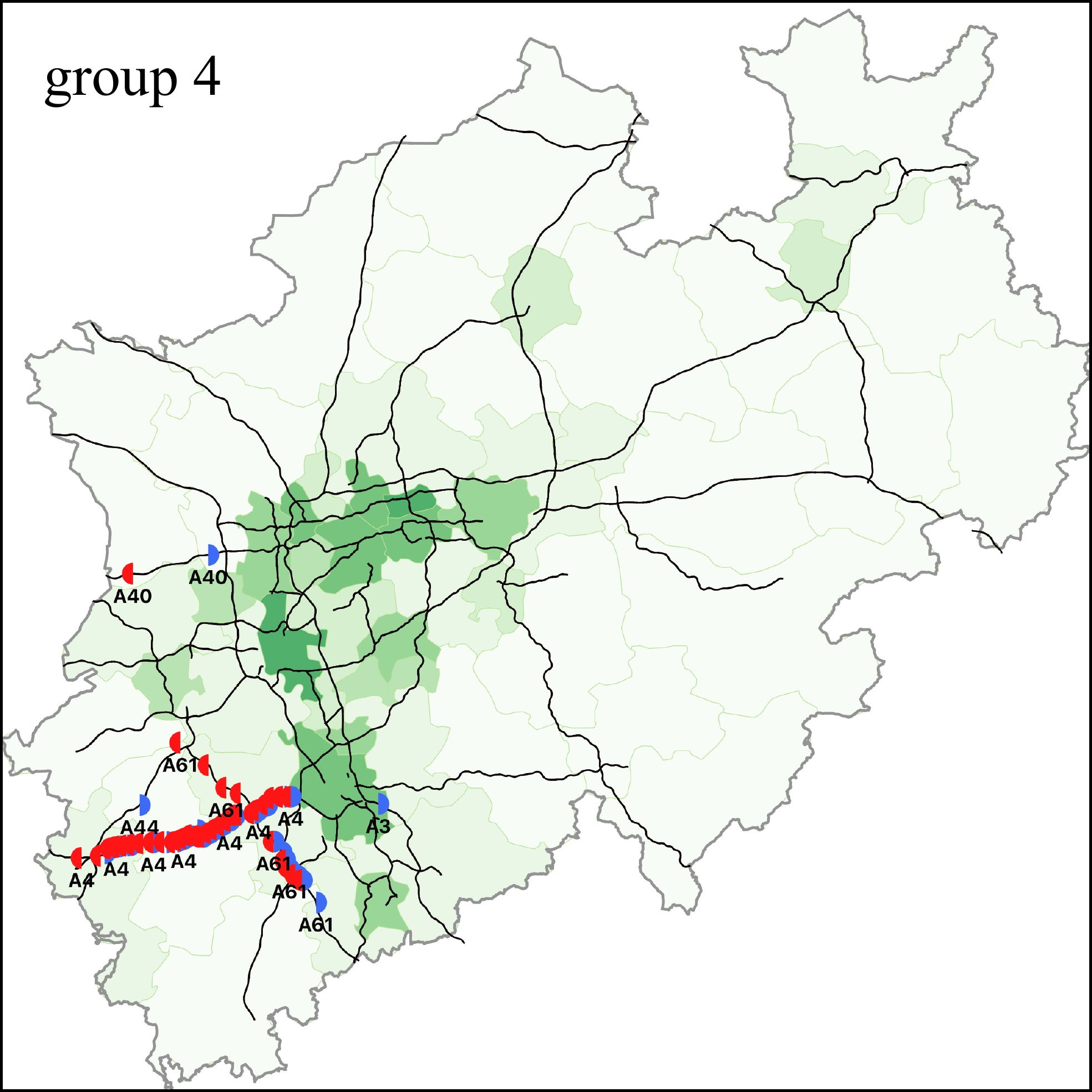} \\ \vspace*{0.07cm}
\floatbox[{\capbeside\thisfloatsetup{capbesideposition={right,center},capbesidewidth=0.465\textwidth}}]{figure}[\FBwidth]
{\caption{Five groups of motorway sections of NRW in Germany for holidays. The green background indicates the population density of districts in NRW. For the information and the data source of the base map, see figure~\ref{fig1}.}
\label{fig9}
}
{\includegraphics[width=0.48\textwidth]{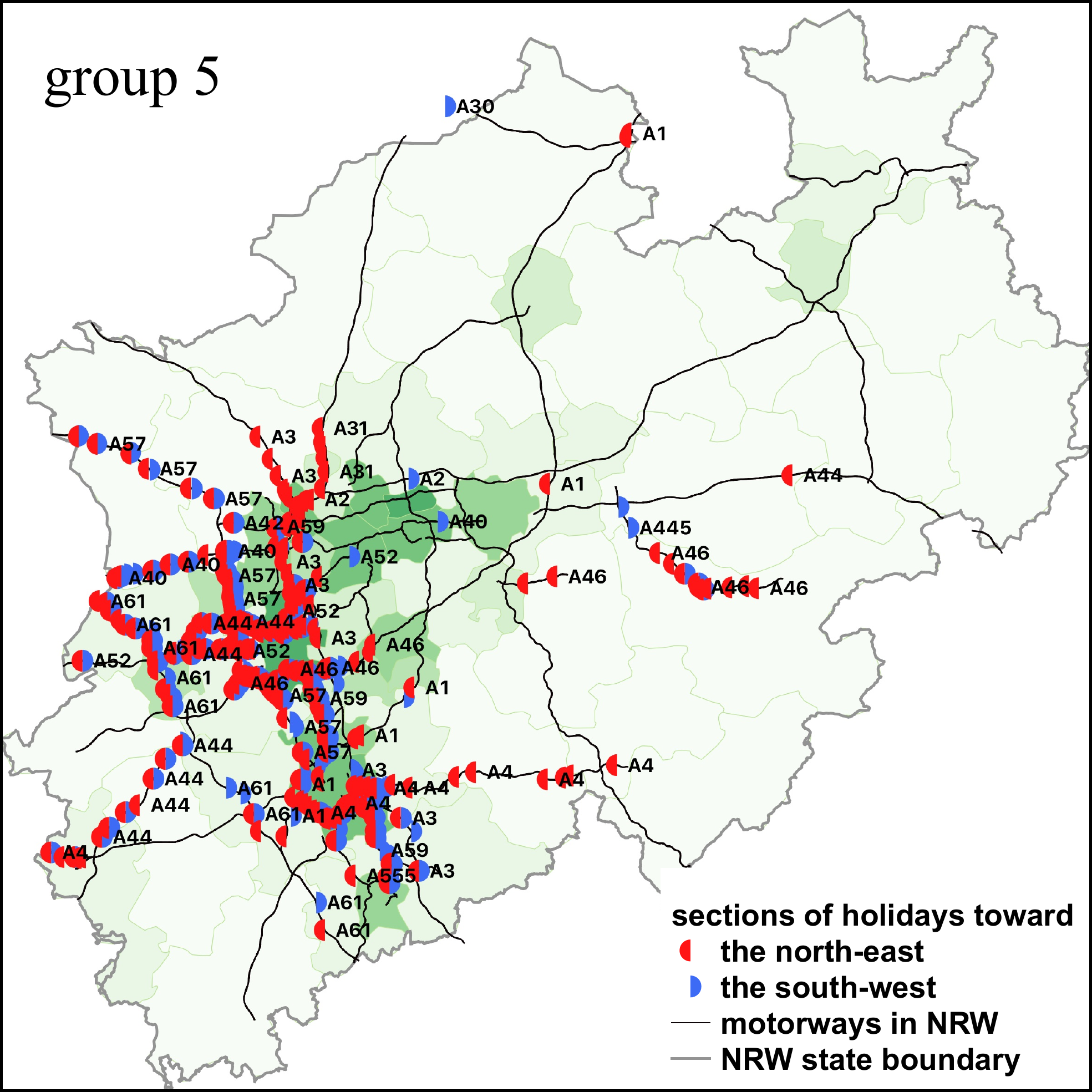}}
\end{center}
\end{figure}

\begin{figure}[tb]
\begin{center}
\includegraphics[width=0.8\textwidth]{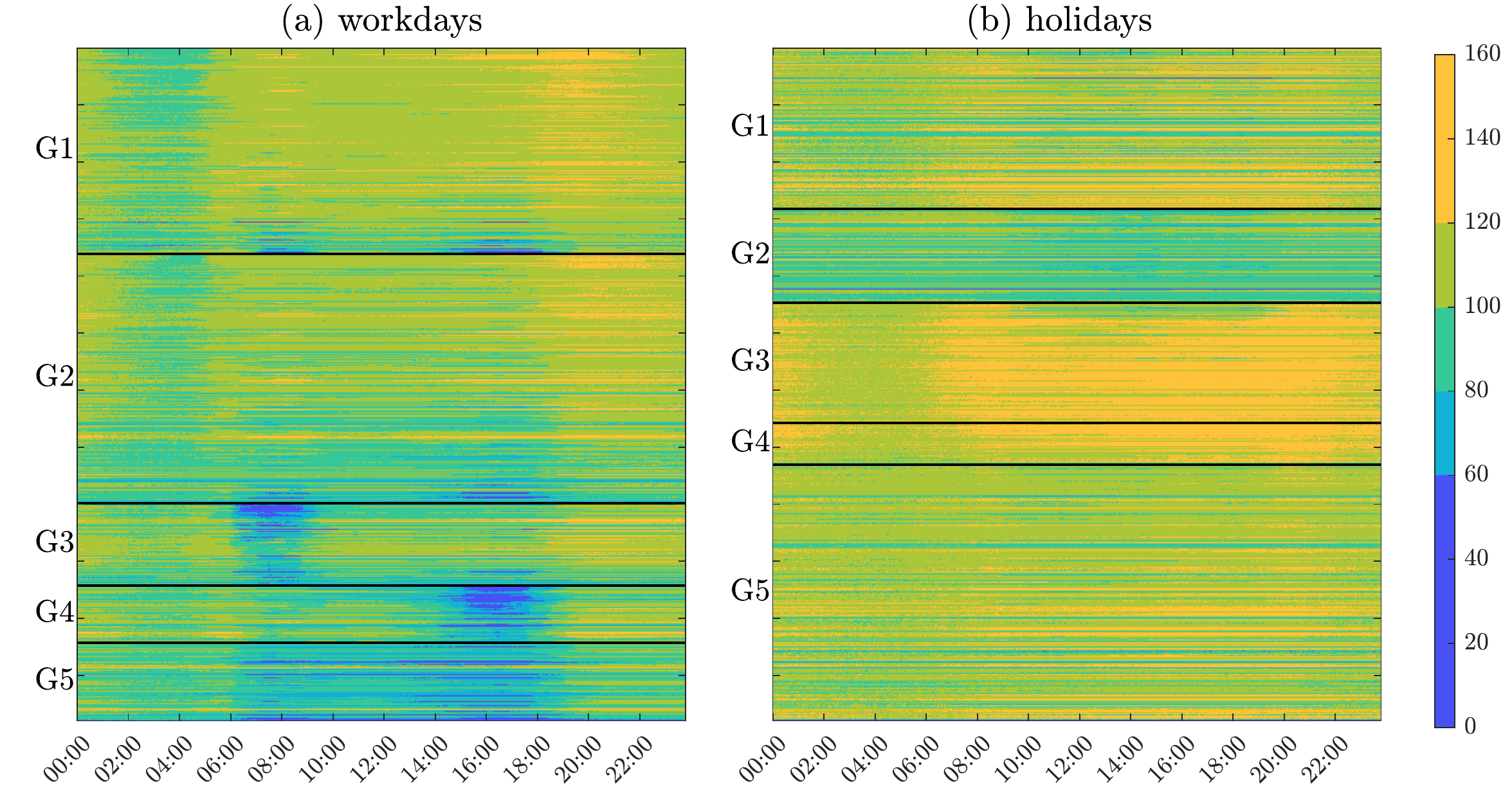}
\caption{Data matrices of dimensions $1179\times 1440$. Each row shows a time series of velocities $v_n(t)$ for section $n$. The data matrices are averaged over all workdays (a) and over all holidays (b), respectively. The rows of data matrices (a) and (b) are ordered corresponding to the reduced-rank correlation matrices in figure~\ref{fig2} (c) and (d), respectively. The black lines distinguish five groups (G1 -- G5). The color indicates the velocity $v_n(t)$ in units of km/h. }
\label{fig10}
\end{center}
\vspace*{-0.5cm}
\end{figure}

Having obtained the correlated groups of sections, we now wish to explore, where on the motorway map the corresponding subdominant effects are active.  We project the sections of each group onto this map, in figure~\ref{fig8} for workdays and in figure~\ref{fig9} for holidays. To better understand the group features, we present the data matrices of velocities in figure~\ref{fig10}. Each row shows a time series of velocities $v_n(t)$ for section $n$.  The data are averaged over all workdays and over all holiday, respectively.

For workdays, the first four groups are strongly correlated. In figure~\ref{fig8}, the sections in groups 1 and 2 almost spread over the whole motorway network, where most sections have very high velocities of $v_n(t)>80$ km/h during a whole day, i.e., between 0:00 and 23:59, as shown in figure~\ref{fig10}. In contrast, the sections in groups 3 and 4 are concentrated in the Rhine-Ruhr metropolitan region with a high population density. The load capacity of each motorway in the Rhine-Ruhr metropolitan region is higher than that in other regions of NRW with a lower population density, especially during rush hours. Along the same motorway, e.g., motorway A3 or A57, in figure~\ref{fig8}, most sections in group 3 are in directions opposite to those in group 4. The difference between the two groups may thus be traced back to the traffic phases during rush hours. We infer from figure~\ref{fig10} that most sections in group 3 are congested ($v_n(t)<60$ km/h) during morning rush hours but free ($v_n(t)>60$ km/h) during afternoon rush hours. For group 4, it is the other way around. Since the commuter traffic flow dominates in rush hours, a majority of commuters go to work passing through the sections in group 3 during morning rush hours and go back home passing through the sections in group 4 during afternoon rush hours. Taking the section directions into account, we can roughly locate the cities where most commuters work. Two of those cities are D\"usseldorf and Cologne. The sections in group 5 are weakly correlated and are scattered over and around the Rhine-Ruhr metropolitan region, see figure~\ref{fig8}. They are slightly congested during day time, i.e., between 6:00 and 18:00.

Groups for holidays present regional features in figure~\ref{fig9} -- middle region for group 1, center region for group 2, right region for group 3, bottom-left region for group 4 and left region for group 5. In particular, groups 2, 3 and 4 belong to strongly correlated groups. In group 2, most sections are concentrated on motorways A40 and A42, where A40 is the most congested motorway in Germany. The velocities on these sections ($80\mathrm{~km/h} <v_n(t)<100$ km/h) are lower than most of those in the other four groups. In contrast, most sections in groups 3 and 4 are far away from the Rhine-Ruhr metropolitan region and their velocities ($v_n(t)>120$ km/h) are remarkably high during day time.

\subsection{Relevant eigenvalues related to geographic distributions of sections}
\label{sec44}

\begin{figure}[tb]
\begin{center}
\begin{overpic}[width=0.45\textwidth]{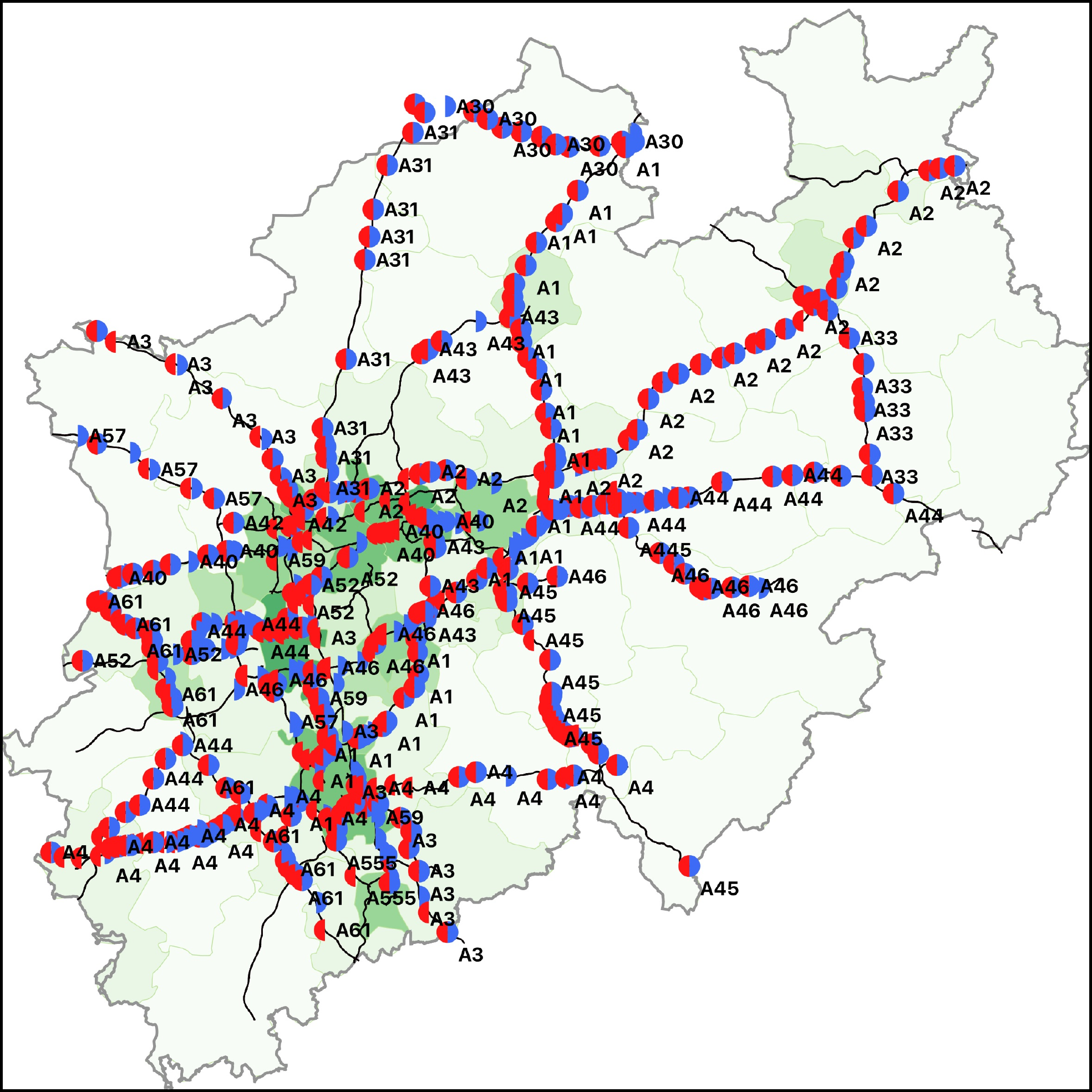} \put (1,93) {\small $\tilde{\lambda}_{N}$, groups 1, 2} \end{overpic}
\vspace*{0.15cm}
\begin{overpic}[width=0.45\textwidth]{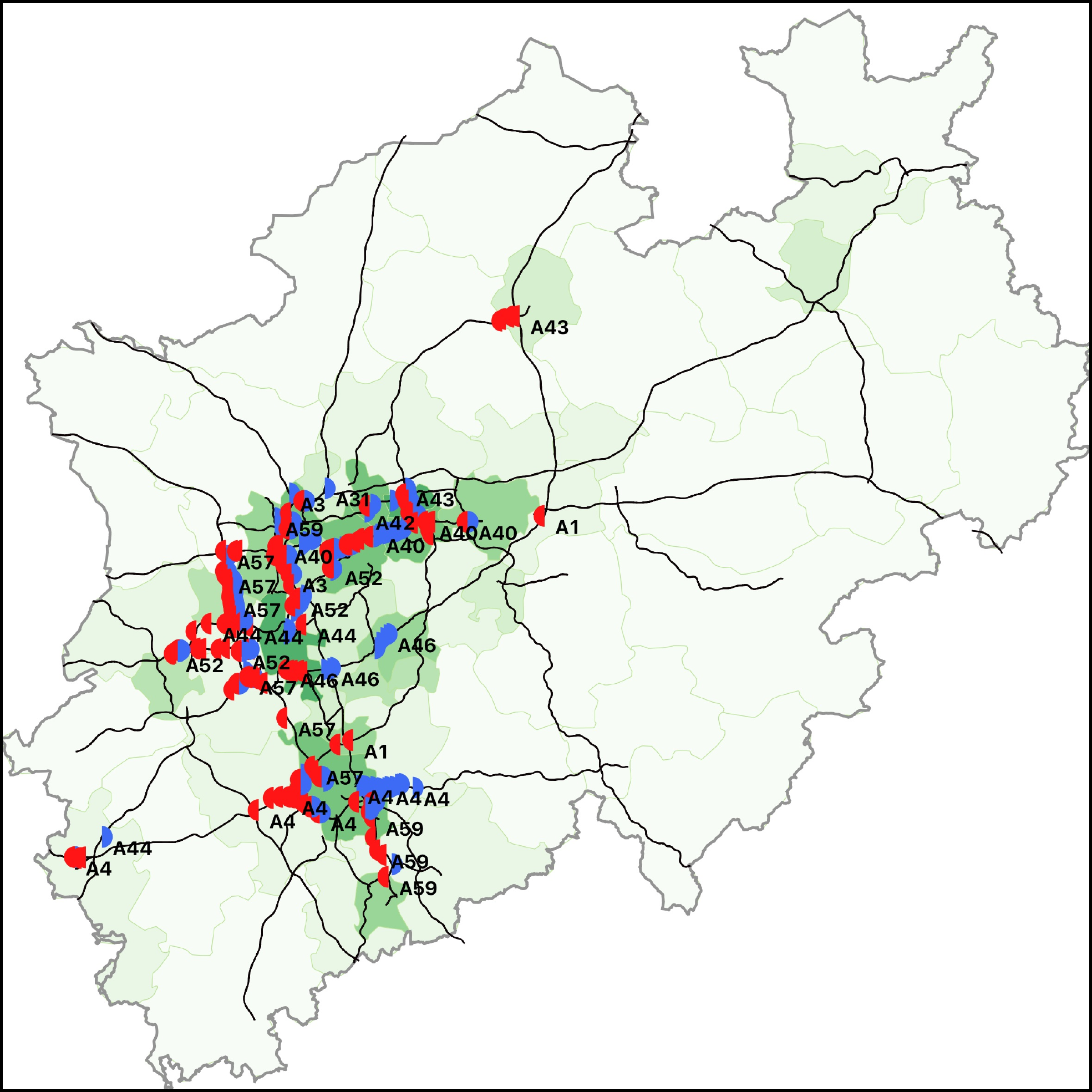} \put (1,93) {\small $\tilde{\lambda}_{N-1}$, groups 3, 4} \end{overpic}
\floatbox[{\capbeside\thisfloatsetup{capbesideposition={right,center},capbesidewidth=0.43\textwidth}}]{figure}[\FBwidth]
{\caption{The geographic distributions of motorway sections of NRW in Germany with the dominant eigenvalues for workdays. The green background indicates the population density of districts in NRW. For the information and the data source of the base map, see figure~\ref{fig1}.}
\label{fig11}}
{\begin{overpic}[width=0.45\textwidth]{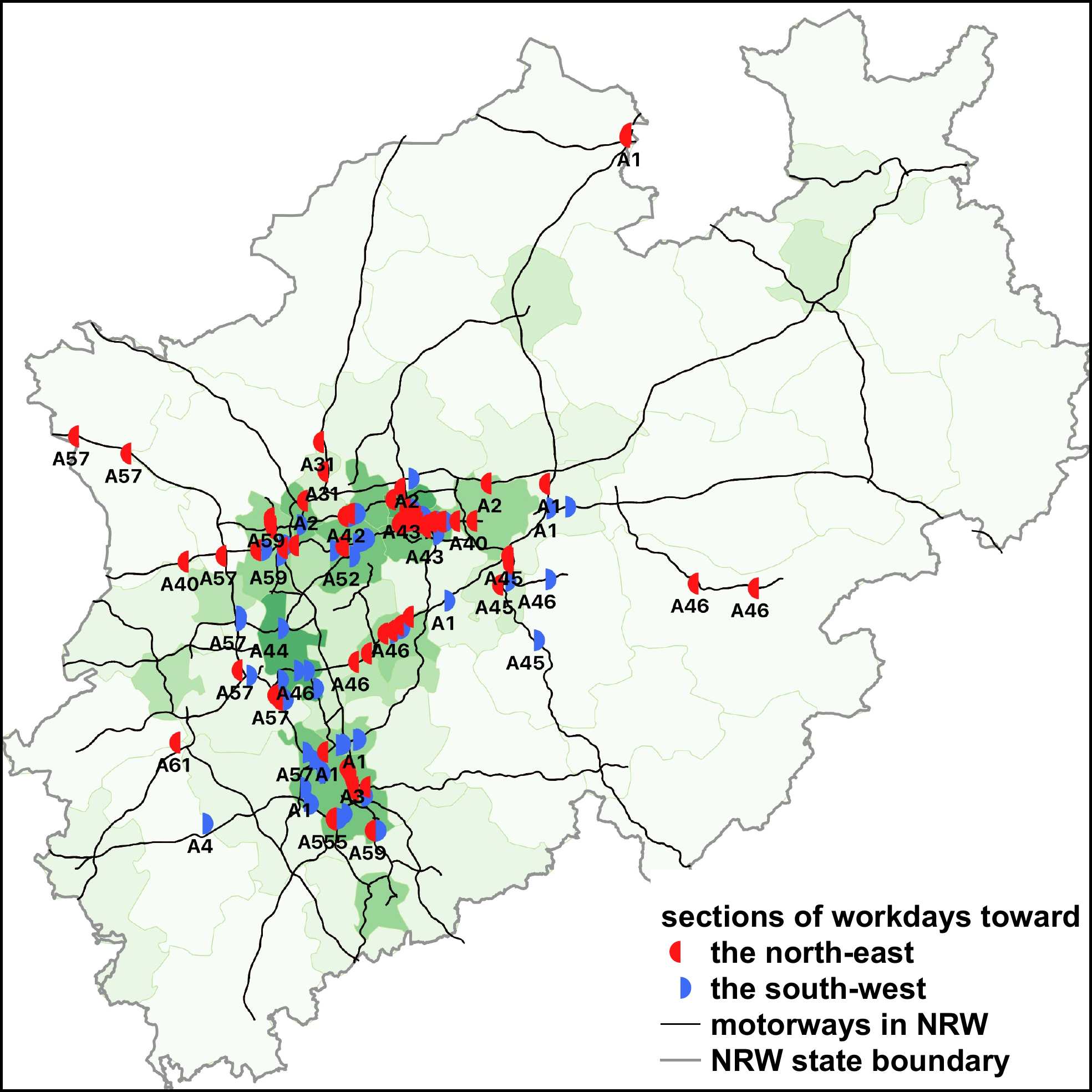} \put (1,93) {\small $\tilde{\lambda}_{N-2}$, group 5} \end{overpic} \vspace*{0.15cm}}
\end{center}
\end{figure}

The spectral features clarify the contributions of the relevant  eigenvalues to each group, while the geographic distributions reveal the geographic location for each group and thus where the subdominant collectivities occur. We now associate the relevant eigenvalues with the geographic distributions of motorway sections for workdays. They are, compared with the case of the holidays, more interesting. A merger using relevant eigenvalues may be possible.  As shown in figure~\ref{fig11}, by combining groups 1 and 2, the sections for the largest eigenvalue $\tilde{\lambda}_{N}$ are distributed on the whole state and most of them are in a free traffic phase at any time of the day. Hence, the largest eigenvalue $\tilde{\lambda}_{N}$ is related to the free traffic phase during a whole day. Regardless of the anti-correlation between groups 3 and 4, their sections for the second largest eigenvalue $\tilde{\lambda}_{N-1}$ are distributed in the Rhine-Ruhr metropolitan region with a high population density. As discussed above, the sections in the two groups are congested during morning or afternoon rush hours due to the commuter traffic flow. Hence, the second largest eigenvalue $\tilde{\lambda}_{N-1}$ is related to the congested traffic phases during rush hours. The sections for the third largest eigenvalue are the ones in group 5 and share the same geographic features with group 5. Hence, the third largest eigenvalue is related to the slightly congested traffic phase during day time.

 \section{Conclusions}
\label{sec5}

We proceeded with our identification of hierarchically ordered collectivities.  Having analyzed the dominant collectivity affecting the entire motorway network in a previous study, we here addressed the subdominant collectivities.  While the former are captured by the largest eigenvalue of the correlation matrix for all motorway sections, the latter are encoded in the next subleading eigenvalues. We succeeded in analyzing them by a proper subtraction of the largest eigenvalue. In contrast to finance, we have to deal with two networks in traffic: a virtual network induced by the correlation coefficients is present in finance and traffic, but in traffic, the ultimate interest is in revealing features of the true motorway network. Put differently, we managed to solve the problem of how to map the information extracted from the virtual network on the true network. To this end, we developed an input-free and seemingly new technique by clustering a (relatively small) number of eigenvectors which uncovers the correlated groups. This step might be viewed as identifying a proper reordering of the basis in the space of the motorway sections. In finance this step is not needed as the assignment of the time series to the industrial sectors is obvious due to elementary economic information.

We analyzed empirical data from 1179 motorway sections of the entire motorway network in NRW, Germany. We identified  five correlated groups. According to the correlation strength, the first four groups for the case of workdays and the middle three groups for the case of holidays were identified as strongly correlated groups. For the case of workdays, the first two groups are governed by the largest eigenvalue of the reduced-rank correlation matrix. Their sections spread to the whole state and most of sections are under a free traffic phase with very high velocities during a whole day. The third and the fourth groups are governed by the second largest eigenvalues. Their sections are concentrated in the Rhine-Ruhr metropolitan region of NRW. Most sections in the third (fourth) group are congested (free) during morning rush hours but free (congested) during afternoon rush hours. The fifth group is a weakly correlated group and governed by the third largest eigenvalue. Its sections are scattered over and around the Rhine-Ruhr metropolitan region and are slightly congested during day time. For the case of holidays, the groups can be separated by regions. All correlated groups correspond to high velocities except the second one whose sections are mainly concentrated on the motorways A40 and A42. The congestion is almost absent on the sections in all groups for the holidays.

The approach developed in this study led to a clear identification and separation of the strongly correlated groups of motorway sections. In particular, the third and the fourth groups identified for workdays are strongly related to the commuter traffic flows during rush hours. The sections in these two groups are more likely to be critical bottlenecks with respect to the load of motorways. From a practical viewpoint, to improve the traffic efficiency, it is better to bypass these sections when determining alternative routes during rush hours. These sections also should be given a priority if any road improvement would be implemented to relieve the load of motorways and enhance the traffic efficiency. From a more conceptual viewpoint, our study of the correlation structure provides completely new information on the network and the dynamics on it. Our approach is also applicable to other correlated and non-stationary complex systems for identifying strongly correlated groups.

\section*{Acknowledgements}
\addcontentsline{toc}{section}{Acknowledgements}
We gratefully acknowledge funding via the grant ``Korrelationen und deren Dynamik in Autobahnnetzen'', Deutsche Forschungsgemeinschaft (DFG, 418382724). We thank Strassen.NRW for providing the empirical traffic data. We also thank Sebastian Gartzke, Anton Josef Heckens, Daniel Waltner and Henrik Bette for fruitful discussions.

\section*{Author contributions}
T.G. and M.S. proposed the research. S.W. and T.G. developed the methods of analysis. S.W. performed all the calculations. S.W. and T.G. wrote the manuscript with the input from M.S. All authors contributed equally to analyzing the results and reviewing the paper.



\end{document}